\documentclass[a4paper,11pt]{article}

\usepackage{jheppub} 
                     
\usepackage[numbers]{natbib}
\usepackage{filecontents}

\usepackage[T1]{fontenc}		

\usepackage{amsmath,amsfonts,amssymb,amstext,mathrsfs}
\usepackage{mathpazo}
\usepackage{epsf,float}

\usepackage{dcolumn}
\usepackage{braket}
\usepackage{color,xcolor}
\usepackage{graphicx}
\usepackage{subfigure}
\usepackage{multirow}
\usepackage{tabularx}
\usepackage{pstricks}
\usepackage[section]{placeins}
\usepackage{booktabs}
\usepackage{array}

\usepackage{hyperref}

\newcommand{\nn}{\nonumber}

\newcommand{\Pom}{\mathbb{P}}

\newcommand{\Reg}{\mathbb{R}}

\usepackage[normalem]{ulem}  

\begin{document}

\title{\boldmath Addendum: Production of $\pi^{+} \pi^{-}$ pairs in diffractive photon-proton and in proton-proton collisions revisited, in particular concerning the Drell-S{\"o}ding contribution}

\author[a,1]{Piotr Lebiedowicz,\note{Corresponding author.}}
\author[b]{Otto Nachtmann}
\author[a,c]{and Antoni Szczurek}

\affiliation[a]{Institute of Nuclear Physics Polish Academy of Sciences, 
Radzikowskiego 152, PL-31342 Krak{\'o}w, Poland}
\affiliation[b]{Institut f\"ur Theoretische Physik, Universit\"at Heidelberg,
Philosophenweg 16, D-69120 Heidelberg, Germany}
\affiliation[c]{Institute of Physics, Faculty of Exact and Technical Sciences, University of Rzesz{\'o}w, 
Pigonia 1, PL-35310 Rzesz{\'o}w, Poland}

\emailAdd{Piotr.Lebiedowicz@ifj.edu.pl}
\emailAdd{O.Nachtmann@thphys.uni-heidelberg.de}
\emailAdd{Antoni.Szczurek@ifj.edu.pl}

\abstract{We update our analyses concerning the non-resonant (Drell-S\"oding)
contribution to the reaction $\gamma^{(*)} p \to \pi^{+} \pi^{-} p$, 
now including general vertex functions
for the coupling of pomeron, $f_{2 \Reg}$, and $\rho_{\Reg}$, to pions.
This takes into account the off-shellness of the pions.
We present a comparison of our tensor-pomeron model,
including our new Drell-S\"oding and $\rho$, $\omega$ vector-meson contributions,
with data for the $\gamma p \to \pi^{+} \pi^{-} p$ reaction
measured by the ZEUS and H1 Collaborations.
We find very good agreement of our theory with the experimental data
for the $\pi^{+} \pi^{-}$ invariant mass ($M_{\pi \pi}$)
and for angular distributions.
Here we restrict our analysis to the region $M_{\pi \pi} \lesssim 1.2$ GeV.
For higher values of $M_{\pi \pi}$ we will have to consider production
of higher-mass mesons.
The purpose of the present article is to show that our tensor-pomeron model
gives a very good fit to the so-called skewing of the $\rho$-resonance
contribution in photoproduction.
In fact, it is an old problem/puzzle to understood the difference
of the shapes of the $M_{\pi \pi}$ distributions in the region
of the $\rho(770)$ resonance in $e^{+} e^{-}$ annihilation
and photoproduction. We think that with our results
we have obtained for this problem a solution which is very satisfactory 
from the Quantum-Field-Theory point of view.
Our results are also relevant for central-exclusive $\pi^{+} \pi^{-}$ production
in ultraperipheral $pp$, A$p$, and AA collisions at the LHC and
for photo- and electroproduction of $\pi^{+} \pi^{-}$ at
future electron-proton/ion colliders, EIC and LHeC.}

\maketitle
\flushbottom

\section{Introduction}
\label{sec:1}

\begin{figure}[tbp]
\centering
(a)\includegraphics[width=7.cm]{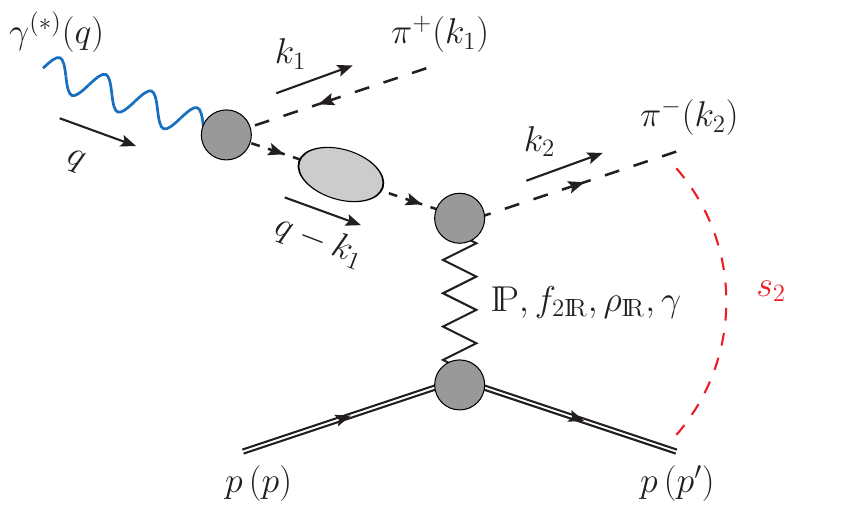}
\hfill
(b)\includegraphics[width=7.cm]{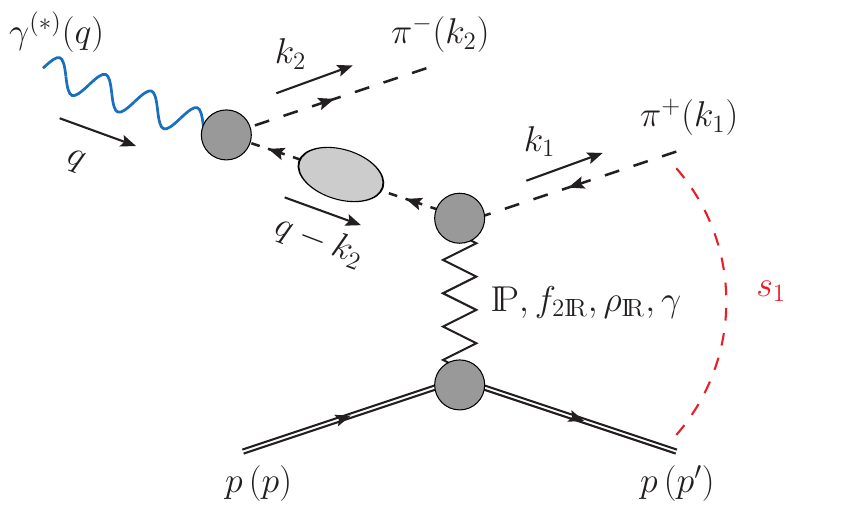}
\hfill
(c)\includegraphics[width=4.6cm]{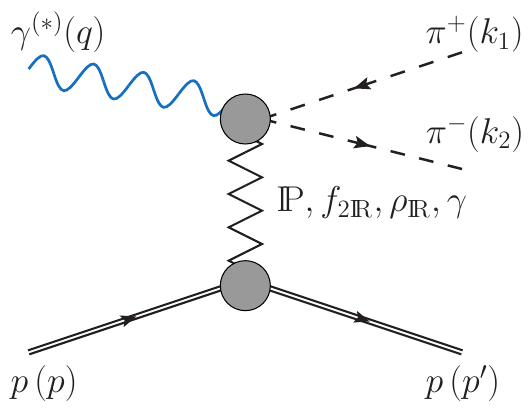}
\caption{\label{fig:100}
Diagrams for non-resonant $\pi^{+} \pi^{-}$ production 
in the $\gamma^{(*)} p \to \pi^{+} \pi^{-} p$ reaction with
$\Pom$, $f_{2 \Reg}$, $\rho_{\Reg}$, and $\gamma$ exchange.}
\end{figure}

In this note we continue our investigation of the production of 
$\pi^{+} \pi^{-}$ pairs in diffractive photon-proton collisions.
In \cite{Lebiedowicz:2025xob} we have given a detailed discussion
of this reaction for real and slightly virtual photons
at high energies, using and extending the results of \cite{Bolz:2014mya}.
In particular, we have presented in \cite{Lebiedowicz:2025xob}
a Quantum-Field-Theory (QFT) treatment of the Drell-S\"oding (DS) term
\cite{Drell:1960zz,Drell:1961zz,Soding:1965nh},
the non-resonant $\pi^{+} \pi^{-}$ production.
Here we consider again the following reaction for real and slightly virtual photons
\begin{eqnarray}
&&\gamma^{(*)}(q, \mu) + p(p, \mathfrak{s}) \to \pi^{+}(k_{1}) + \pi^{-}(k_{2}) + p(p',\mathfrak{s'})\,, \nn \\
&&k = k_{1} + k_{2}\,.
\label{1.1}
\end{eqnarray}
Here $q$, $p$, $p'$, $k_{1}$, $k_{2}$ are the momenta of the particles,
$k$ is the four-momentum of the $\pi^{+}\pi^{-}$ system,
$\mathfrak{s}$, $\mathfrak{s'}$ are the spin indices of the protons,
and $\mu$ is the vector index of the photon;
see (2.1) of \cite{Lebiedowicz:2025xob}.
We require $0 \leqslant -q^{2} \lesssim 0.5~{\rm GeV}^{2}$.
The diagrams for the DS term are shown in figure~1 of \cite{Lebiedowicz:2025xob}.
To make the present article self contained we reproduce this figure here
(see figure~\ref{fig:100}).

Note that in the diagrams (a) and (b) of figure~\ref{fig:100}
we have the full $\gamma \pi \pi$ vertex functions and the full
pion propagators.
Using the generalized Ward identity for the pions
we showed in \cite{Lebiedowicz:2025xob} that for real photons
the product of these two factors in the diagrams of 
figures~\ref{fig:100}(a) and \ref{fig:100}(b) is fixed
up to an irrelevant gauge term;
see (A.27) and (A.28) of \cite{Lebiedowicz:2025xob}.
For small virtuality of the photons we made the reasonable ansatz
for this product, (A.32) in \cite{Lebiedowicz:2025xob}.
Let us now consider the pomeron couplings
in figures~\ref{fig:100}(a) and \ref{fig:100}(b).
We use the framework of the tensor-pomeron model \cite{Ewerz:2013kda}.
The $\Pom pp$ coupling is here for on-shell protons
and is well known, as is the effective pomeron propagator;
see (3.10), (3.11), (3.43), and (3.44) of \cite{Ewerz:2013kda}.
Also the $\Pom \pi \pi$ vertex function 
is well known for \textit{on-shell} pions;
see (3.45) and (3.46) of \cite{Ewerz:2013kda}.
But in figures~\ref{fig:100}(a) and \ref{fig:100}(b)
we have the $\Pom \pi \pi$ vertices with one pion being off shell.
In \cite{Lebiedowicz:2025xob} we took also for this case the same
$\Pom \pi \pi$ vertex as for both pions on shell.
In the present paper we generalize our model by allowing
for off-shell effects in the $\Pom \pi \pi$, $f_{2 \Reg} \pi \pi$,
and $\rho_{\Reg} \pi \pi$ vertices occurring in figures~\ref{fig:100}(a) and \ref{fig:100}(b).
As we did in \cite{Lebiedowicz:2025xob} we shall then use
the gauge-invariance constraints to calculate the expressions
for the diagram of figure~\ref{fig:100}(c).

Our paper is organized as follows.

In section~\ref{sec:2}, we derive the new amplitudes for the DS term.
In section~\ref{sec:3}, we present a comparison of our tensor-pomeron model
with data for the $\gamma p \to \pi^{+} \pi^{-} p$ reaction
measured by the ZEUS \cite{ZEUS:1997rof} and H1 \cite{H1:2020lzc} Collaborations.
In section~\ref{sec:4} we summarize our findings.
In appendix~\ref{sec:A} we compare our results to those presented
by J.~Pumplin in \cite{Pumplin:1970kp}.

\section{Updated Drell-S\"oding amplitudes taking into account off-shell pion effects}
\label{sec:2}

As mentioned in the introduction we shall now calculate
the DS amplitudes taking into account off-shell effects
in the $\Pom \pi \pi$, $f_{2 \Reg} \pi \pi$,
and $\rho_{\Reg} \pi \pi$ effective vertices.
For the general discussion of these vertices
we refer to \cite{Lebiedowicz:2025xob,Bolz:2014mya,Ewerz:2013kda}.

We start with the $\Pom \pi \pi$ vertex.

First, we note that from $C$ invariance 
we have the same vertex function for $\Pom \pi^{-} \pi^{-}$
and $\Pom \pi^{+} \pi^{+}$; see eq.~(A.38) of \cite{Lebiedowicz:2025xob}.
Keeping the tensor part of this vertex as given there 
we can only generalize the form factor in (A.38) of \cite{Lebiedowicz:2025xob} by making the following replacement:
\begin{eqnarray}
F_{M}[(\tilde{k}'-\tilde{k})^{2}] \to 
F[(\tilde{k}'-\tilde{k})^{2},\tilde{k}'^{2}-m_{\pi}^{2},\tilde{k}^{2}-m_{\pi}^{2}]\,.
\label{2.100}
\end{eqnarray}
In this way we get
\begin{align}
i\Gamma_{\mu \nu}^{(\Pom \pi \pi)}(\tilde{k}',\tilde{k})
=& -i 2 \beta_{\Pom \pi \pi} 
F[(\tilde{k}'-\tilde{k})^{2},\tilde{k}'^{2}-m_{\pi}^{2},\tilde{k}^{2}-m_{\pi}^{2}]\nn\\
& \times
\left[ (\tilde{k}' + \tilde{k})_{\mu} 
       (\tilde{k}' + \tilde{k})_{\nu}
- \frac{1}{4} g_{\mu \nu} (\tilde{k}' + \tilde{k})^{2} \right]\,.
\label{A1}
\end{align}
The coupling constant $\beta_{\Pom \pi \pi}$ 
is as in (A.38) of \cite{Lebiedowicz:2025xob}.
The crossing relations of QFT require
\begin{align}
\Gamma_{\mu \nu}^{(\Pom \pi \pi)}(-\tilde{k},-\tilde{k}')
= \Gamma_{\mu \nu}^{(\Pom \pi \pi)}(\tilde{k}',\tilde{k})\,,
\label{A2}
\end{align}
which imply
\begin{align}
F[(\tilde{k}'-\tilde{k})^{2},\tilde{k}'^{2}-m_{\pi}^{2},\tilde{k}^{2}-m_{\pi}^{2}]
=
F[(\tilde{k}'-\tilde{k})^{2},\tilde{k}^{2}-m_{\pi}^{2},\tilde{k}'^{2}-m_{\pi}^{2}]\,.
\label{A3}
\end{align}
According to the Landau conditions these functions are analytic for
\begin{align}
\tilde{k}^{2} - m_{\pi}^{2} < 8 m_{\pi}^{2}\,, \quad
\tilde{k}'^{2} - m_{\pi}^{2} < 8 m_{\pi}^{2}\,, \quad
(\tilde{k}'-\tilde{k})^{2} < 4 m_{\pi}^{2}\,.
\label{A4}
\end{align}
As normalisation condition we take
\begin{align}
F(0,0,0)=1\,.
\label{A5}
\end{align}

Now we turn to section~2.1 of \cite{Lebiedowicz:2025xob}
which is devoted to the Drell-S{\"o}ding amplitudes for pomeron exchange.
In the expressions (2.12) and (2.13) of \cite{Lebiedowicz:2025xob}
we have to replace in ${\cal F}_{\Pom \pi p}(2 \nu_{1,2}, t)$
the form factor $F_{M}(t)$,
see (A.52) of \cite{Lebiedowicz:2025xob}, by $F(.)$ from (\ref{A3}).
In this way we get new amplitudes
denoted by ${\cal N}^{(a,b)}_{\mu, \mathfrak{s'},\mathfrak{s}}$ as follows.

We define for $k^{2} = m_{\pi}^{2}$ and 
$t_{\pi,i} = (q - k_{i})^{2}$, $i = 1,2$,
\begin{eqnarray}
G[t, t_{\pi,i} - m_{\pi}^{2}] = 
F[t, 0, t_{\pi,i} - m_{\pi}^{2}] (F_{M}(t))^{-1}\,.
\label{A6}
\end{eqnarray}
We get then
\begin{eqnarray}
{\cal N}^{(0,a)}_{\mathfrak{s'},\mathfrak{s}}(k_{2},p',q-k_{1},p)|_{\Pom}&=&
G(t,t_{\pi,1}-m_{\pi}^{2})
{\cal M}^{(0,a)}_{\mathfrak{s'},\mathfrak{s}}(k_{2},p',q-k_{1},p)|_{\Pom}\,,\label{A8}\\
{\cal N}^{(a)}_{\mu, \mathfrak{s'},\mathfrak{s}}(k_{1},k_{2},p',q,p)|_{\Pom}&=&
G(t,t_{\pi,1}-m_{\pi}^{2})
{\cal M}^{(a)}_{\mu, \mathfrak{s'},\mathfrak{s}}(k_{1},k_{2},p',q,p)|_{\Pom}\,, 
\label{A9}
\end{eqnarray}
and
\begin{eqnarray}
{\cal N}^{(0,b)}_{\mathfrak{s'},\mathfrak{s}}(k_{1},p',q-k_{2},p)|_{\Pom}&=&
G(t,t_{\pi,2}-m_{\pi}^{2})
{\cal M}^{(0,b)}_{\mathfrak{s'},\mathfrak{s}}(k_{1},p',q-k_{2},p)|_{\Pom}\,,
\label{A10}\\
{\cal N}^{(b)}_{\mu, \mathfrak{s'},\mathfrak{s}}(k_{1},k_{2},p',q,p)|_{\Pom} &=&
G(t,t_{\pi,2}-m_{\pi}^{2})
{\cal M}^{(b)}_{\mu, \mathfrak{s'},\mathfrak{s}}(k_{1},k_{2},p',q,p)|_{\Pom}\,. 
\label{A11}
\end{eqnarray}
Here and in the following we denote by ${\cal M}^{(0,a)}_{\mathfrak{s'},\mathfrak{s}}$,
${\cal M}^{(a)}_{\mu,\mathfrak{s'},\mathfrak{s}}$ etc. 
the DS amplitudes of \cite{Lebiedowicz:2025xob}
and by ${\cal N}^{(0,a)}_{\mathfrak{s'},\mathfrak{s}}$,
${\cal N}^{(a)}_{\mu,\mathfrak{s'},\mathfrak{s}}$ etc.
the corresponding new amplitudes which take into account the generalized form factors 
in the $\Pom \pi \pi$, $f_{2 \Reg} \pi \pi$, 
and $\rho_{\Reg} \pi \pi$ vertices.

The gauge-invariance relation (2.11) of \cite{Lebiedowicz:2025xob} 
reads now for $\Pom$ exchange
\begin{eqnarray}
q^{\mu} {\cal N}^{(c)}_{\mu, \mathfrak{s'},\mathfrak{s}}(k_{1},k_{2},p',q,p)|_{\Pom}
= e 
\left[
{\cal N}^{(0,a)}_{\mathfrak{s'},\mathfrak{s}}(k_{2},p',q-k_{1},p)|_{\Pom}
-
{\cal N}^{(0,b)}_{\mathfrak{s'},\mathfrak{s}}(k_{1},p',q-k_{2},p)|_{\Pom}
\right]. \qquad
\label{A12}
\end{eqnarray}
Inserting (\ref{A8}) and (\ref{A10}) in (\ref{A12}) we find:
\begin{eqnarray}
q^{\mu} {\cal N}^{(c)}_{\mu, \mathfrak{s'},\mathfrak{s}}(k_{1},k_{2},p',q,p)|_{\Pom}
&=&
\frac{1}{2} \left[ 
G(t,t_{\pi,1}-m_{\pi}^{2}) +
G(t,t_{\pi,2}-m_{\pi}^{2}) \right]
e
\left[
{\cal M}^{(0,a)}_{\mathfrak{s'},\mathfrak{s}}|_{\Pom}
-
{\cal M}^{(0,b)}_{\mathfrak{s'},\mathfrak{s}}|_{\Pom}
\right] \nn \\
&+&
\frac{1}{2} \left[ 
G(t,t_{\pi,1}-m_{\pi}^{2}) -
G(t,t_{\pi,2}-m_{\pi}^{2}) \right]
e
\left[
{\cal M}^{(0,a)}_{\mathfrak{s'},\mathfrak{s}}|_{\Pom}
+
{\cal M}^{(0,b)}_{\mathfrak{s'},\mathfrak{s}}|_{\Pom}
\right].\nn \\
\label{A13}
\end{eqnarray}
Now we use the results of (2.11), (2.19) and (2.23) of \cite{Lebiedowicz:2025xob}
to get from (\ref{A13})
\begin{eqnarray}
q^{\mu} {\cal N}^{(c)}_{\mu, \mathfrak{s'},\mathfrak{s}}(k_{1},k_{2},p',q,p)|_{\Pom}
&=&
\frac{1}{2} \left[ 
G(t,t_{\pi,1}-m_{\pi}^{2}) +
G(t,t_{\pi,2}-m_{\pi}^{2}) \right]
q^{\mu} {\cal M}^{(c)}_{\mu, \mathfrak{s'},\mathfrak{s}}(k_{1},k_{2},p',q,p)|_{\Pom} \nn \\
&+&
q^{\mu} (k_{2} - k_{1})_{\mu}\,
H(t,t_{\pi,1}-m_{\pi}^{2},t_{\pi,2}-m_{\pi}^{2})\,
e \left[
{\cal M}^{(0,a)}_{\mathfrak{s'},\mathfrak{s}}|_{\Pom}
+
{\cal M}^{(0,b)}_{\mathfrak{s'},\mathfrak{s}}|_{\Pom}
\right],\nn \\
\label{A14}
\end{eqnarray}
where
\begin{align}
H(t,t_{\pi,1}-m_{\pi}^{2},t_{\pi,2}-m_{\pi}^{2})
=\frac{G(t,t_{\pi,1}-m_{\pi}^{2}) - G(t,t_{\pi,2}-m_{\pi}^{2})}{(t_{\pi,1} - m_{\pi}^{2}) - (t_{\pi,2} - m_{\pi}^{2})}\,.
\label{A15}
\end{align}
Due to (\ref{A4}) for the functions of (\ref{A3}) also $H$
has the analyticity region (\ref{A4})
and no singularity for $k_{1} = k_{2}$
which implies $t_{\pi,1} = t_{\pi,2}$.

The simplest solution of (\ref{A14}) for 
${\cal N}^{(c)}_{\mu, \mathfrak{s'},\mathfrak{s}}|_{\Pom}$ is,
therefore,
\begin{eqnarray}
{\cal N}^{(c)}_{\mu, \mathfrak{s'},\mathfrak{s}}(k_{1},k_{2},p',q,p)|_{\Pom} &=&
\frac{1}{2} \left[ 
G(t,t_{\pi,1}-m_{\pi}^{2}) +
G(t,t_{\pi,2}-m_{\pi}^{2}) \right]
{\cal M}^{(c)}_{\mu, \mathfrak{s'},\mathfrak{s}}(k_{1},k_{2},p',q,p)|_{\Pom}\nn  
\\
&&+ \,(k_{2} - k_{1})_{\mu}\,
H(t,t_{\pi,1}-m_{\pi}^{2},t_{\pi,2}-m_{\pi}^{2}) \nn  \\
&&\times
e \left[
{\cal M}^{(0,a)}_{\mathfrak{s'},\mathfrak{s}}(k_{2},p',q-k_{1},p)|_{\Pom}
+ {\cal M}^{(0,b)}_{\mathfrak{s'},\mathfrak{s}}(k_{1},p',q-k_{2},p)|_{\Pom}
\right]. \nn \\
\label{A16}
\end{eqnarray}

Our result for the DS pomeron-exchange term considering a general
$\Pom \pi \pi$ vertex function of the type of eq.~(\ref{A1}) is,
therefore,
\begin{eqnarray}
{\cal N}^{(\rm DS)}_{\mu, \mathfrak{s'},\mathfrak{s}}|_{\Pom} =
{\cal N}^{(a)}_{\mu, \mathfrak{s'},\mathfrak{s}}|_{\Pom} +
{\cal N}^{(b)}_{\mu, \mathfrak{s'},\mathfrak{s}}|_{\Pom} + 
{\cal N}^{(c)}_{\mu, \mathfrak{s'},\mathfrak{s}}|_{\Pom}\,.
\label{A17}
\end{eqnarray}
See eqs.~(\ref{A9}), (\ref{A11}), (\ref{A16}), and figure~\ref{fig:100}.

Analogously to the $\Pom$-exchange term we can treat the $f_{2 \Reg}$
and $\rho_{\Reg}$ exchange terms.
For this we replace in (A.39) and (A.40) of \cite{Lebiedowicz:2025xob} 
the factor $F_{M}[(\tilde{k}'-\tilde{k})^{2}]$ 
by the function $F(.)$ as in (\ref{2.100}).
In principle there could be different functions $F(.)$
for $\Pom$, $f_{2 \Reg}$, and $\rho_{\Reg}$ exchanges,
but they all must satisfy conditions (\ref{A4}).
However, here we assume the same function for all these exchanges.
Then the results (\ref{A9}), (\ref{A11}), (\ref{A16}), and (\ref{A17})
apply also for the $f_{2 \Reg}$ and $\rho_{\Reg}$ exchanges.
In this way we get with 
the results of sections~2.1, 2.2.3 and 2.2.4 of \cite{Lebiedowicz:2025xob}
for the amplitudes
${\cal M}^{(\rm DS) (a,b,c)}_{\mu, \mathfrak{s'},\mathfrak{s}}$:
\begin{eqnarray}
{\cal N}^{(\rm DS)}_{\mu, \mathfrak{s'},\mathfrak{s}} =
{\cal N}^{(\rm DS)}_{\mu, \mathfrak{s'},\mathfrak{s}}|_{\Pom} +
{\cal N}^{(\rm DS)}_{\mu, \mathfrak{s'},\mathfrak{s}}|_{f_{2 \Reg}} + 
{\cal N}^{(\rm DS)}_{\mu, \mathfrak{s'},\mathfrak{s}}|_{\rho_{\Reg}}\,.
\label{2.17a}
\end{eqnarray}

In the calculation we assume the function $G(t,t_{\pi,i}-m_{\pi}^{2})$
in the form
%
\begin{eqnarray}
&&G(t,t_{\pi,i}-m_{\pi}^{2}) = F_{M}(t) \, F_{\pi}(t_{\pi,i})\,, \nn \\
&&i = 1,2\,,
\label{A19}
\end{eqnarray}
with
\begin{eqnarray}
F_{M}(t) &=& \frac{\Lambda^{2}}{\Lambda^{2} + t}\,, 
\label{A20}\\
F_{\pi}(t_{\pi,i}) &=& \exp\left( \frac{t_{\pi,i} - m_{\pi}^{2}}{\Lambda_{\pi}^{2}}\right)\,, \quad i = 1,2\,.
\label{A21}
\end{eqnarray}
We take
$\Lambda^{2} = 0.75$~GeV$^{2}$ (see figure~13 of \cite{Lebiedowicz:2025xob})
and  
$\Lambda_{\pi} = 0.8$~GeV (see figures~\ref{fig:1}--\ref{fig:5} and figure~\ref{fig:DS_comparison} below).
In addition, in the bottom panels of Fig.~\ref{fig:1}, 
we show the results for $\Lambda_{\pi} = 1.0$~GeV.

In the following
\begin{eqnarray}
W_{\gamma p} = \sqrt{(p+q)^{2}} \,,
\label{A22}
\end{eqnarray}
is the c.m. energy of the photon-proton collision.

\section{Results for the $\gamma p \to \pi^{+} \pi^{-} p$ reaction and a comparison with data from HERA}
\label{sec:3}

In this section, we present a comparison of our tensor-pomeron model
with data for the $\gamma p \to \pi^{+} \pi^{-} p$ reaction
measured by the ZEUS \cite{ZEUS:1997rof} and H1 \cite{H1:2020lzc} Collaborations.

The results presented here correspond to the complete tensor-pomeron model
including also reggeon ($f_{2 \Reg}$, $\rho_{\Reg}$) exchanges.
In the calculations of the $\rho(770)$ contribution
we use the parameter set~A
of the $\Pom \rho \rho$ and $f_{2 \Reg} \rho \rho$ coupling constants; 
see (B.2) of \cite{Lebiedowicz:2025xob}
and figure 12 therein.
The form factors $\tilde{F}^{(V)}(k^{2})$ ($V = \rho, \omega$)
occurring in the $\Pom VV$ and $f_{2 \Reg} VV$ vertices
are taken of the standard form
\begin{equation}
\tilde{F}^{(V)}(k^{2}) = 
\left[ 1 + \frac{k^{2}(k^{2} - m_{V}^{2})}{\Lambda_{V}^{4}} \right]^{-n_{V}}\,,
\label{ff_V}
\end{equation}
$V = \rho, \omega$, $\Lambda_{\rho} = \Lambda_{\omega}$,
$n_{\rho} = n_{\omega}$;
see (2.29) of \cite{Lebiedowicz:2025xob}
and (B.82)--(B.88) of \cite{Bolz:2014mya}.
The production of the $f_{2}(1270)$ meson 
decaying to $\pi^{+} \pi^{-}$ is predicted to be very small
and will be neglected in the following;
see figures~3 and 4 of \cite{Bolz:2014mya}.
For the pomeron trajectory, we use the parameters
$\epsilon_{\Pom} = 0.0808$
and $\alpha_{\Pom}' = 0.25$~GeV$^{-2}$, 
see eq.~(A.45) of \cite{Lebiedowicz:2025xob}.
The secondary reggeon trajectories are parametrized
with the intercepts 
$\alpha_{f_{2 \Reg}}(0) = \alpha_{\rho_{\Reg}}(0) = 0.5475$
and the slopes $\alpha_{f_{2 \Reg}}' = \alpha_{\rho_{\Reg}}' = 0.9$~GeV$^{-2}$,
in accordance with eqs.~(A.47) and (A.49) of \cite{Lebiedowicz:2025xob}.

\begin{figure}[tbp]
\centering
\includegraphics[width=0.48\textwidth]{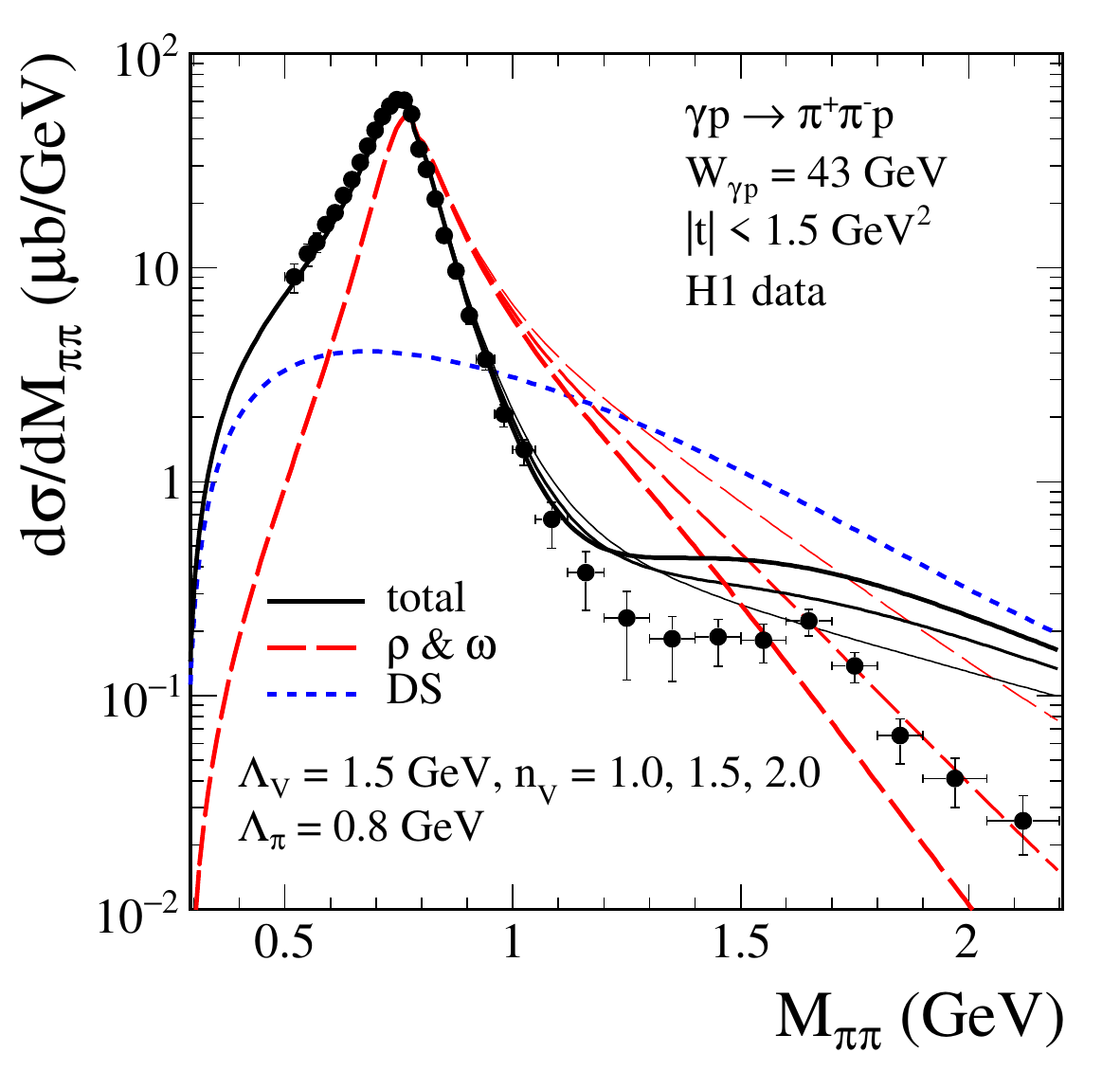}
\includegraphics[width=0.48\textwidth]{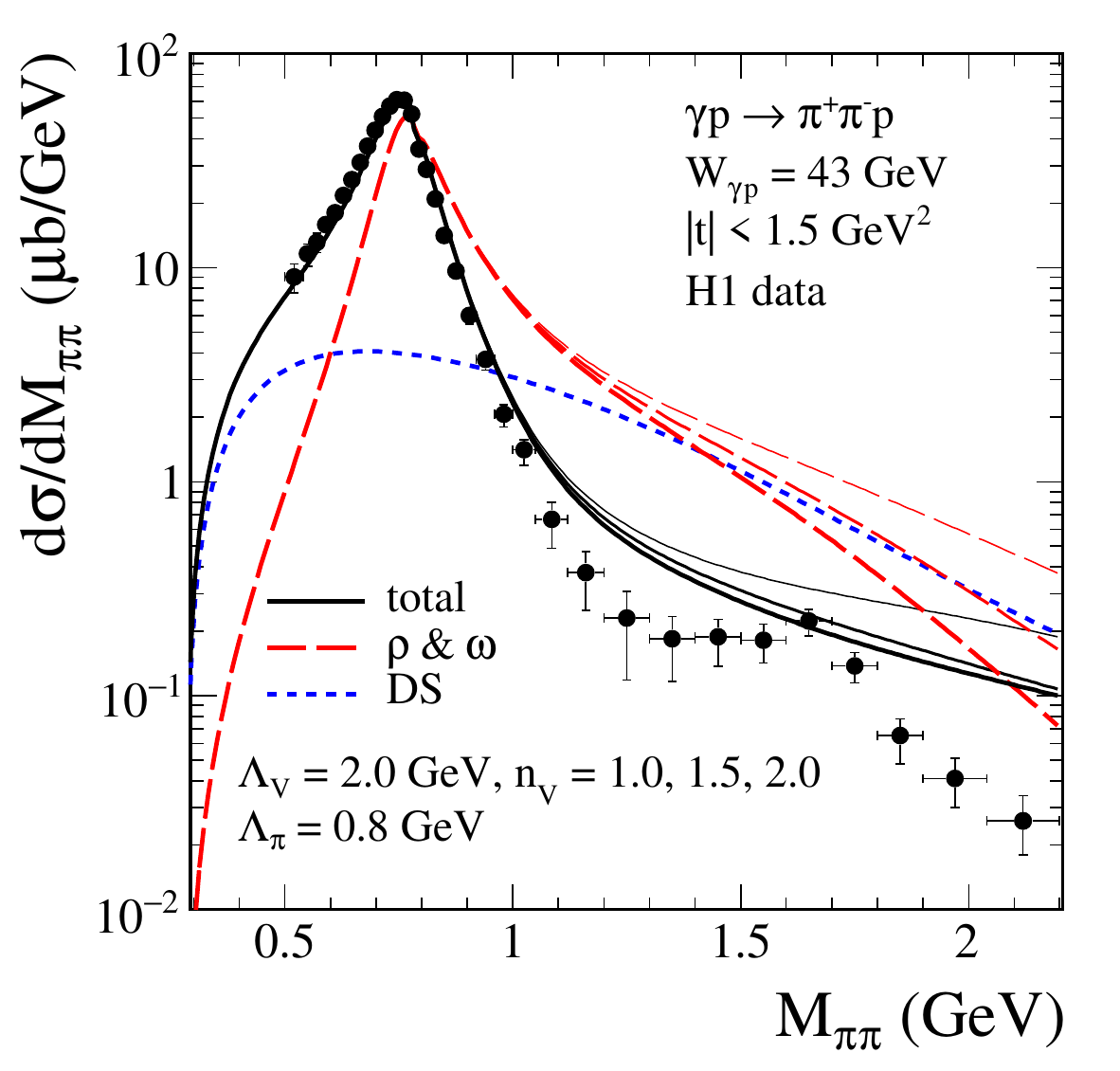}
\includegraphics[width=0.48\textwidth]{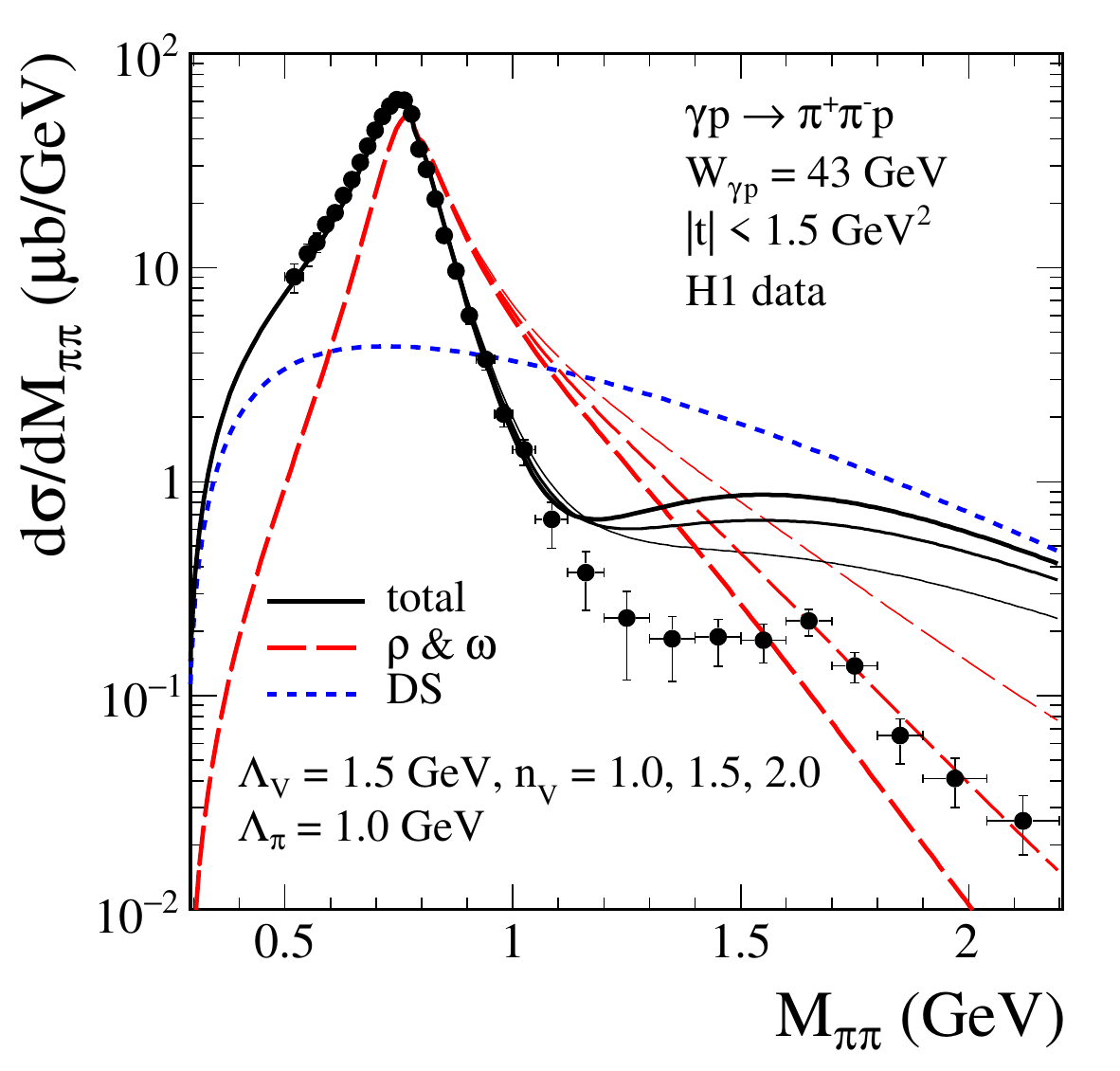}
\includegraphics[width=0.48\textwidth]{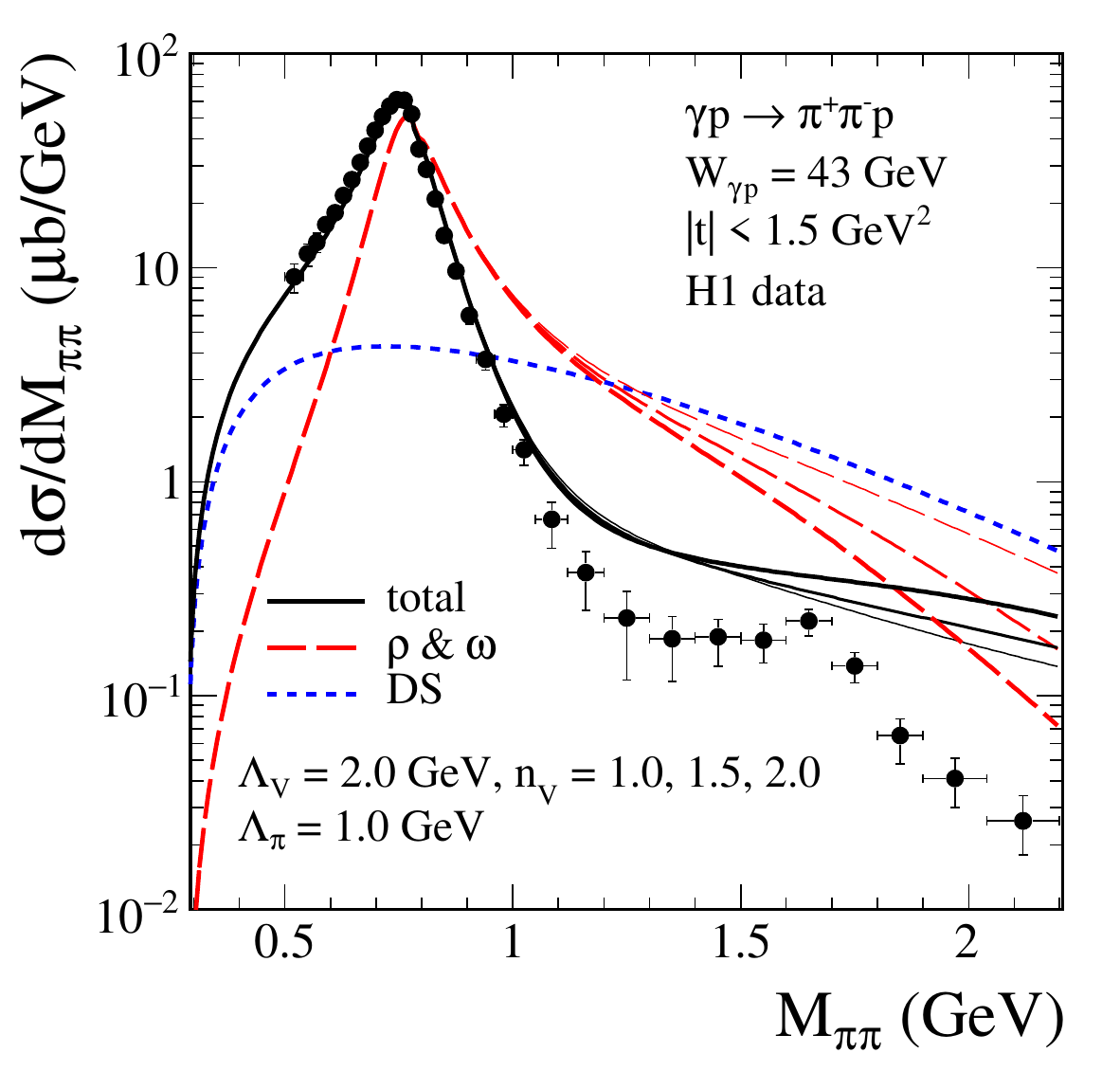}
\caption{\label{fig:1}
The invariant mass $M_{\pi\pi}$ distribution 
measured by the H1 Collaboration \cite{H1:2020lzc}
for $20 < W_{\gamma p} < 80$~GeV 
($\langle W_{\gamma p} \rangle = 43$~GeV) and $|t| < 1.5$~GeV$^{2}$
compared to the model results calculated 
for fixed $W_{\gamma p} = 43$~GeV.
The statistical and systematic
uncertainties are shown.
Presented are complete results (total) and results 
for the resonant ($\rho\, \& \,\omega$) 
and the non-resonant Drell-S\"oding (DS) contributions.
For the DS term in (\ref{A21}),
we take $\Lambda_{\pi} = 0.8$~GeV (top panels)
or $\Lambda_{\pi} = 1.0$~GeV (bottom panels).
The resonant term is calculated 
with $\tilde{F}^{(V)}(k^{2})$ for $V = \rho, \omega$
as in (\ref{ff_V}),
with the $\Lambda_{V}$ and $n_{V}$ parameters 
indicated in the legend.
The left panels present the results for 
$\Lambda_{V} = 1.5$~GeV and $n_{V} = 1.0, 1.5, 2.0$
(see the red long-dashed and black solid lines from thin to thick).
Similarly, the right panels show the results 
for $\Lambda_{V} = 2.0$~GeV and $n_{V} = 1.0, 1.5, 2.0$ (from thin to thick).
No $\rho(1450)$, $\rho_{3}(1690)$, and $\rho(1700)$ 
resonances are taken into account here.}
\end{figure}

In figure~\ref{fig:1} we show the distribution of the $\pi^{+} \pi^{-}$
invariant mass calculated at fixed photon-proton centre-of-mass energy
$W_{\gamma p} = 43$~GeV and for $|t| < 1.5$~GeV$^{2}$.
We show the model results for the resonant ($\rho^{0}$ and $\omega$)
and the DS contributions.
Our results are compared with the H1 data from \cite{H1:2020lzc}.
We explicitly show the curves for different values of the form factor parameters 
[$\Lambda_{\pi}$ in (\ref{A21}), 
$\Lambda_{V}$ and $n_V$ in (\ref{ff_V})]
which roughly illustrate the model uncertainty.
A precise estimation of the theoretical uncertainties 
is rather difficult due to strong interference effects. 
Our model, with the selected parametrization of the form factors, 
does not describe the region $M_{\pi \pi} \gtrsim 1.1$ GeV.
However, in this region, other mechanisms
(for instance, the production of $\rho(1450)$, $\rho_{3}(1690)$, and $\rho(1700)$ resonances)
are possible, and fitting our current model to the H1 data is not justified.
Restricting our analysis to 
the low-$M_{\pi \pi}$ region in the following figures, 
we choose $\Lambda_{\pi} = 0.8$~GeV, $\Lambda_{V} = 2.0$~GeV, 
and $n_{V} = 2.0$.

\begin{figure}[tbp]
\centering
\includegraphics[width=0.48\textwidth]{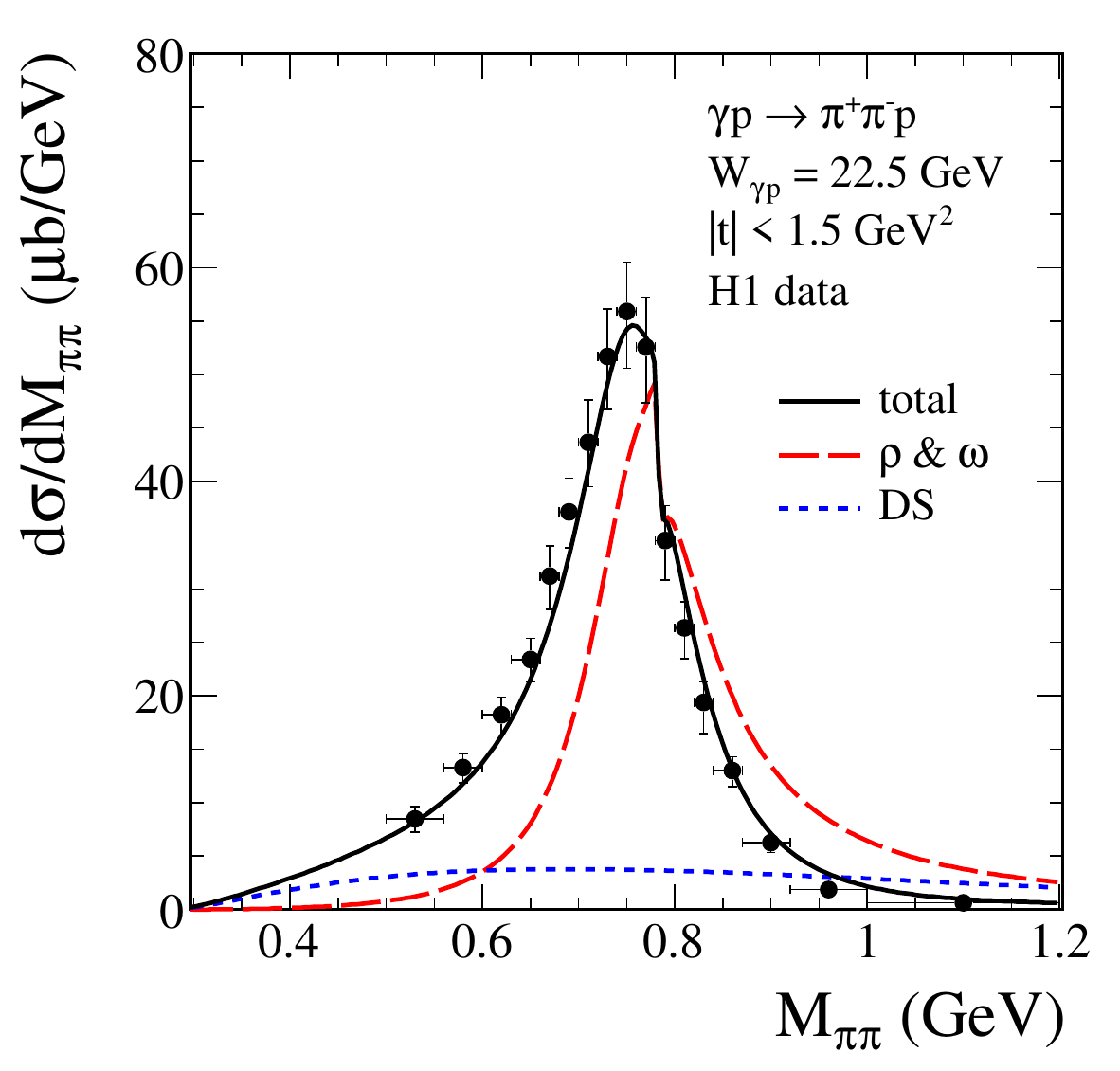}
\includegraphics[width=0.48\textwidth]{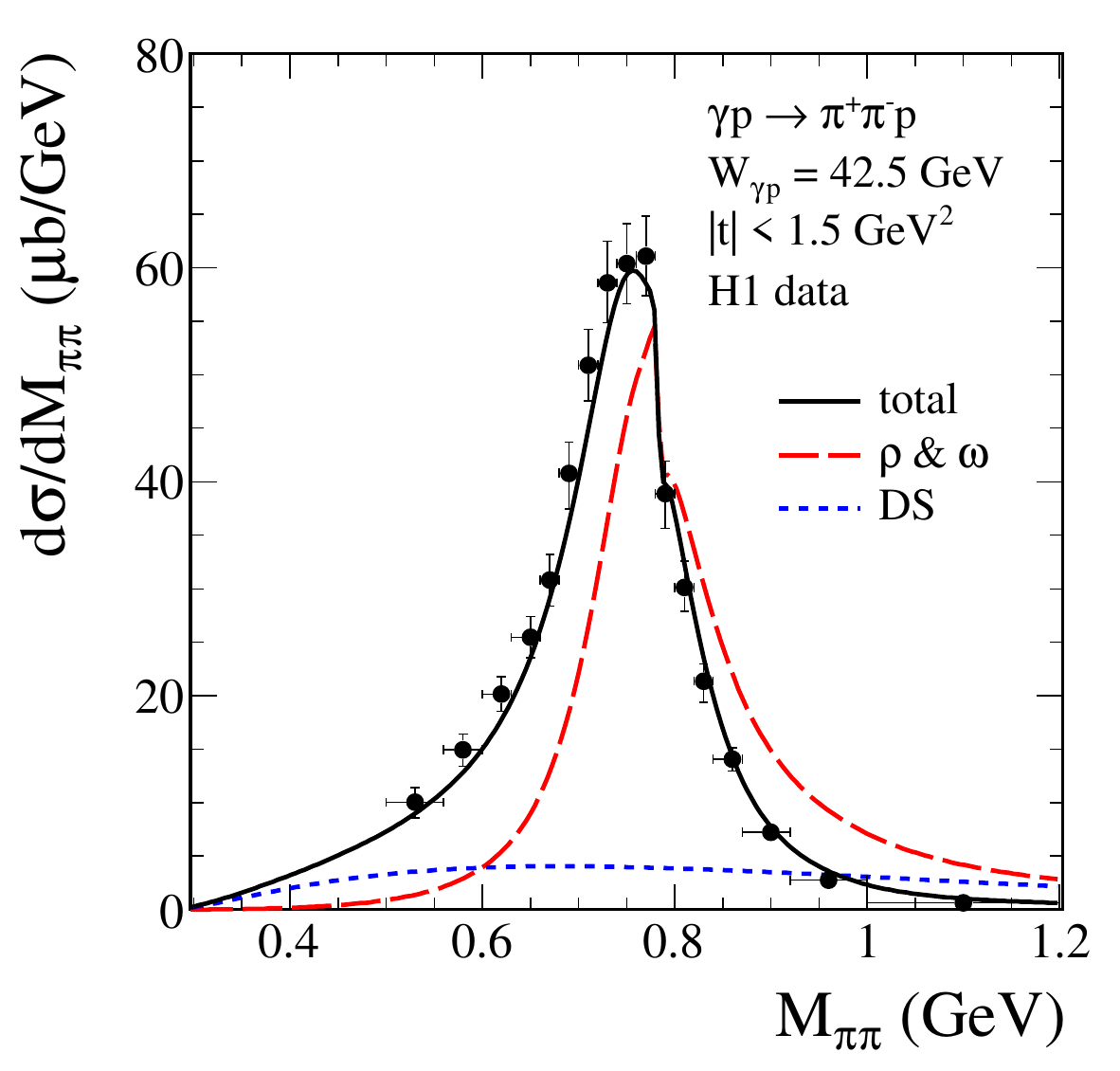}
\includegraphics[width=0.48\textwidth]{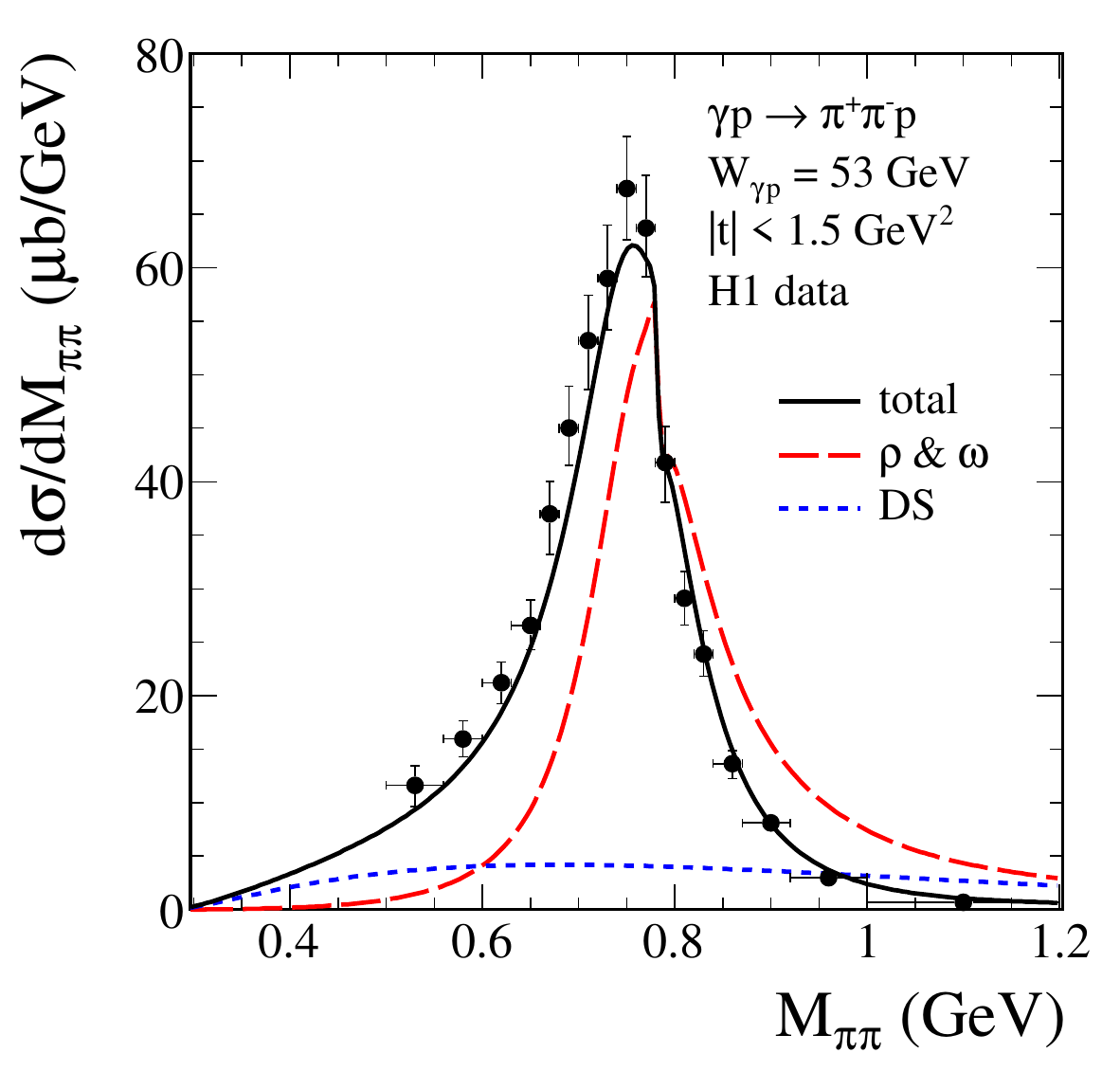}
\includegraphics[width=0.48\textwidth]{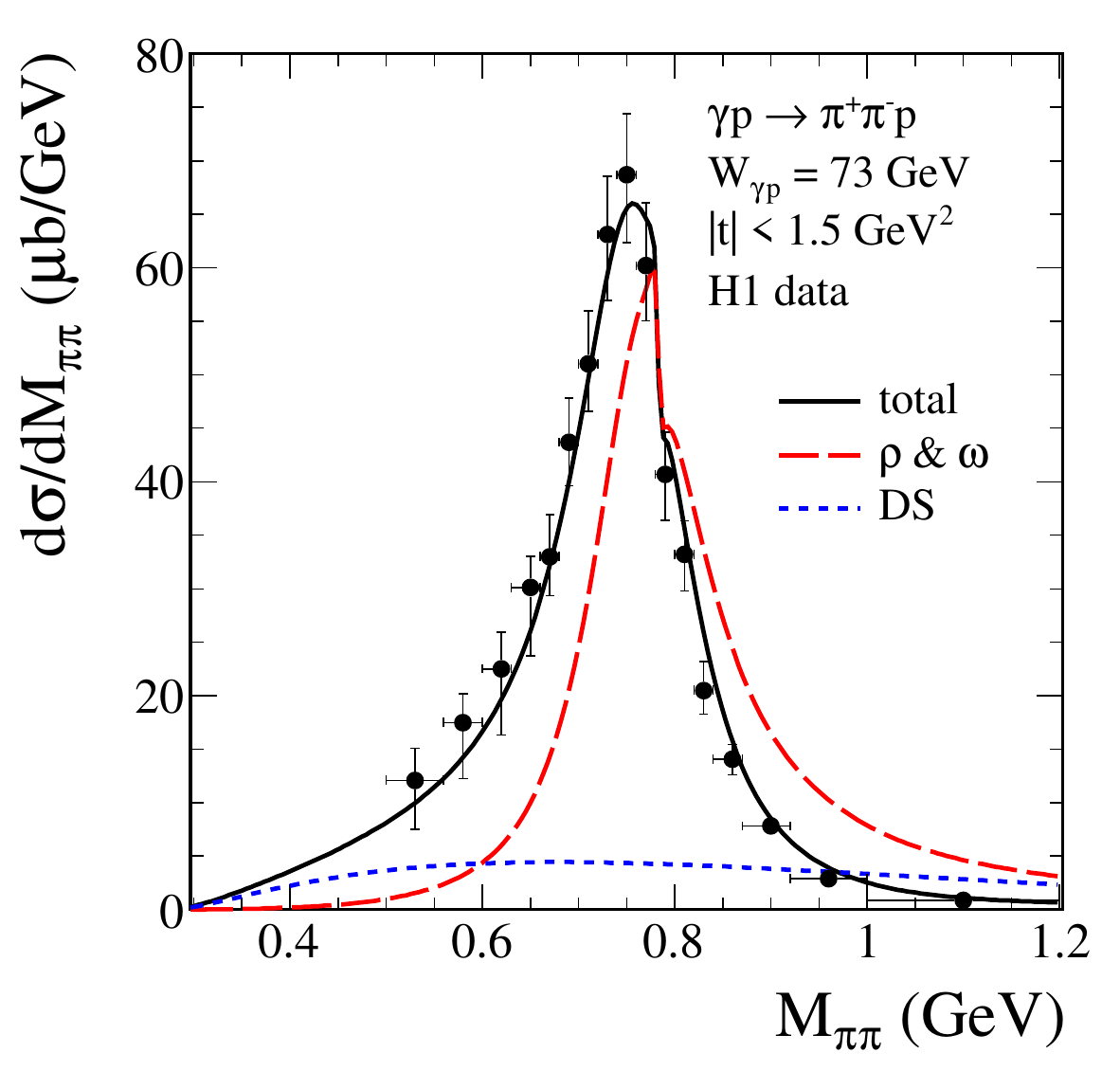}
\caption{\label{fig:2}
The differential distributions 
for the reaction $\gamma p \to \pi^{+} \pi^{-} p$
for fixed $W_{\gamma p}$.
The experimental data of the H1 Collaboration are from \cite{H1:2020lzc}
and were measured in the ranges $20 < W_{\gamma p} < 25$~GeV,
$40 < W_{\gamma p} < 45$~GeV,
$50 < W_{\gamma p} < 56$~GeV,
and $66 < W_{\gamma p} < 80$~GeV.
The statistical and systematic
uncertainties are shown.
The meaning of the lines is the same as in figure~\ref{fig:1}.
In the calculations we take $\Lambda_{\pi} = 0.8$~GeV, $\Lambda_{V} = 2.0$~GeV, 
and $n_{V} = 2.0$.}
\end{figure}

\begin{figure}[tbp]
\centering
\includegraphics[width=0.48\textwidth]{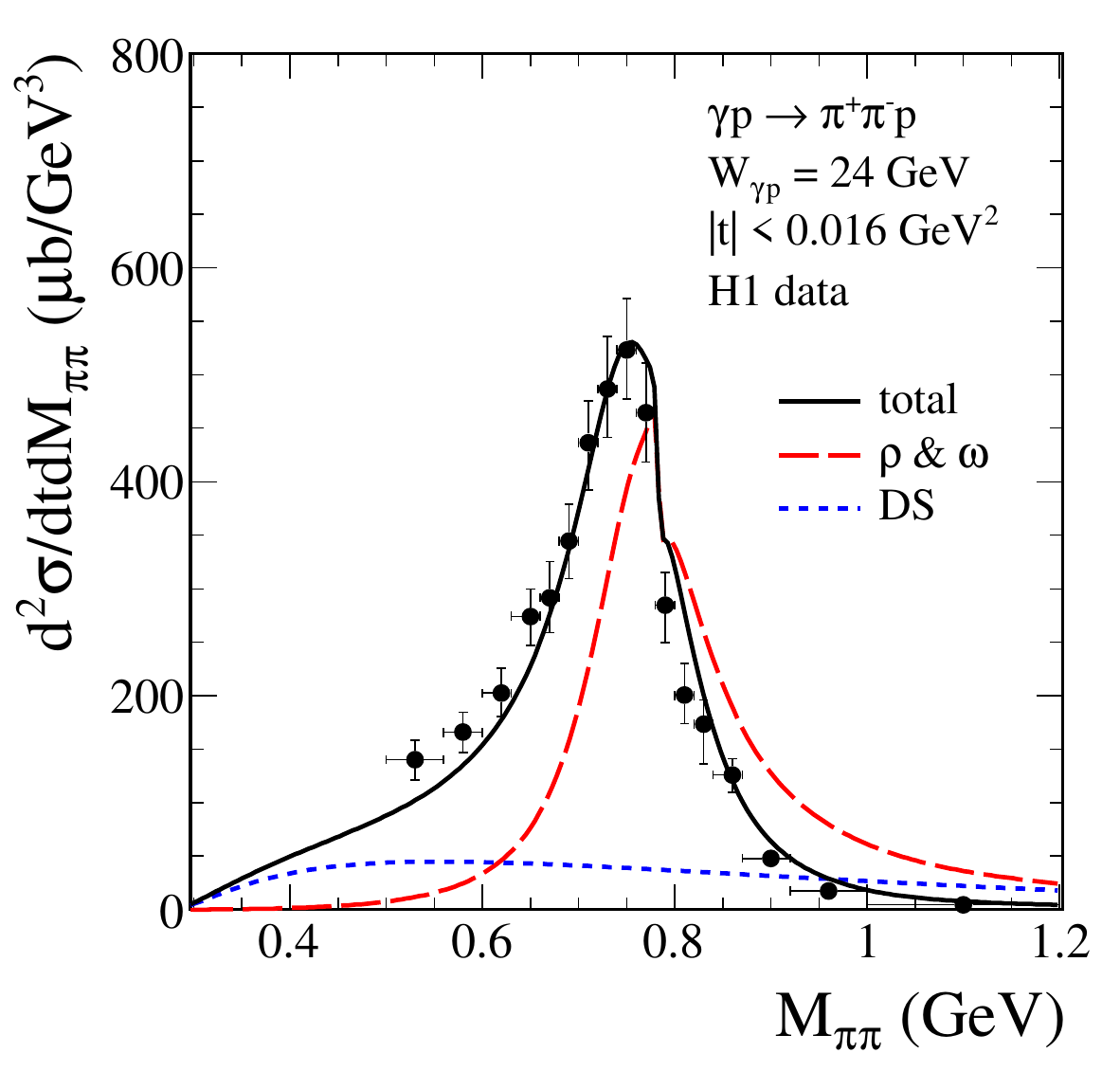}
\includegraphics[width=0.48\textwidth]{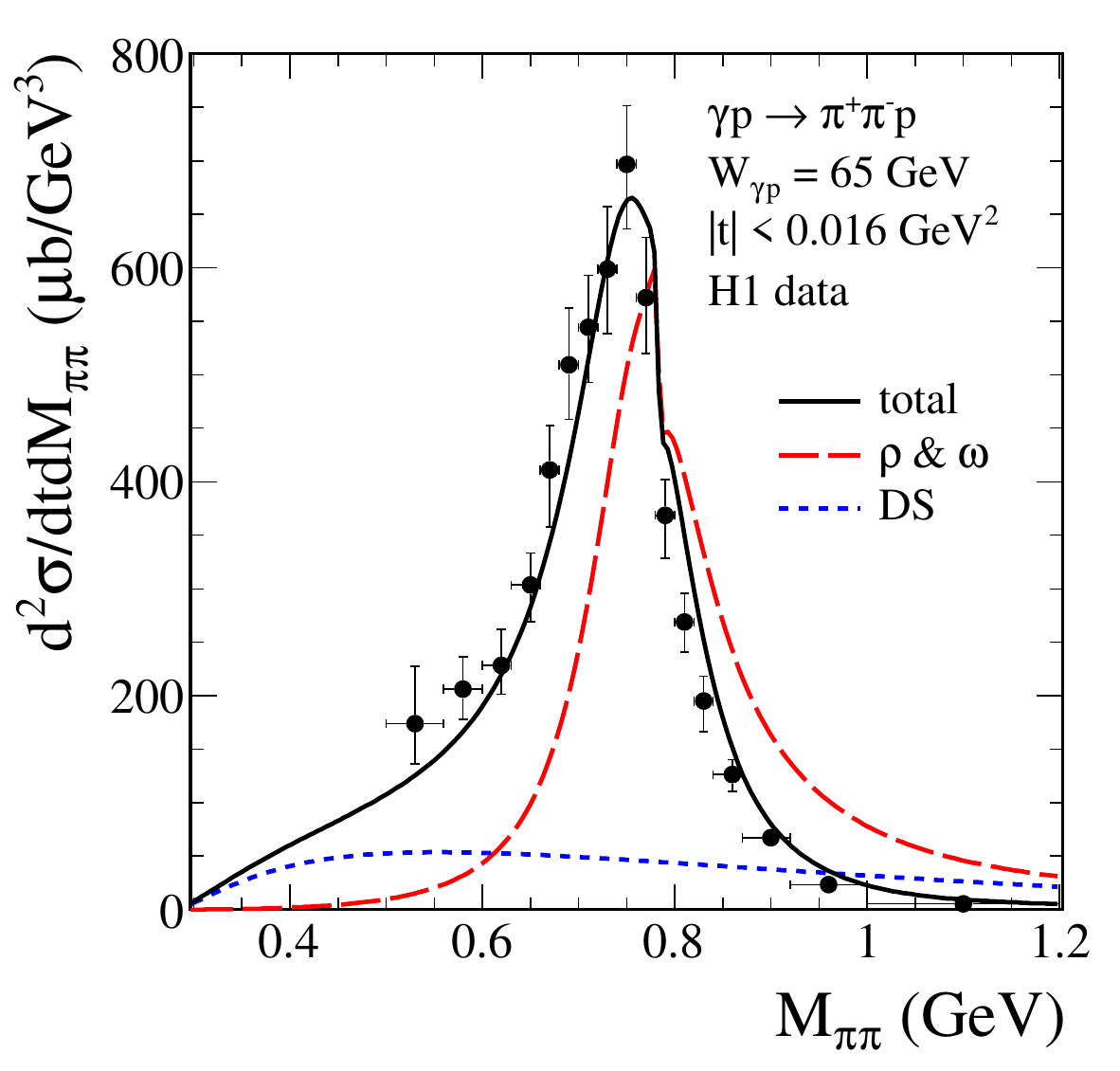}
\includegraphics[width=0.48\textwidth]{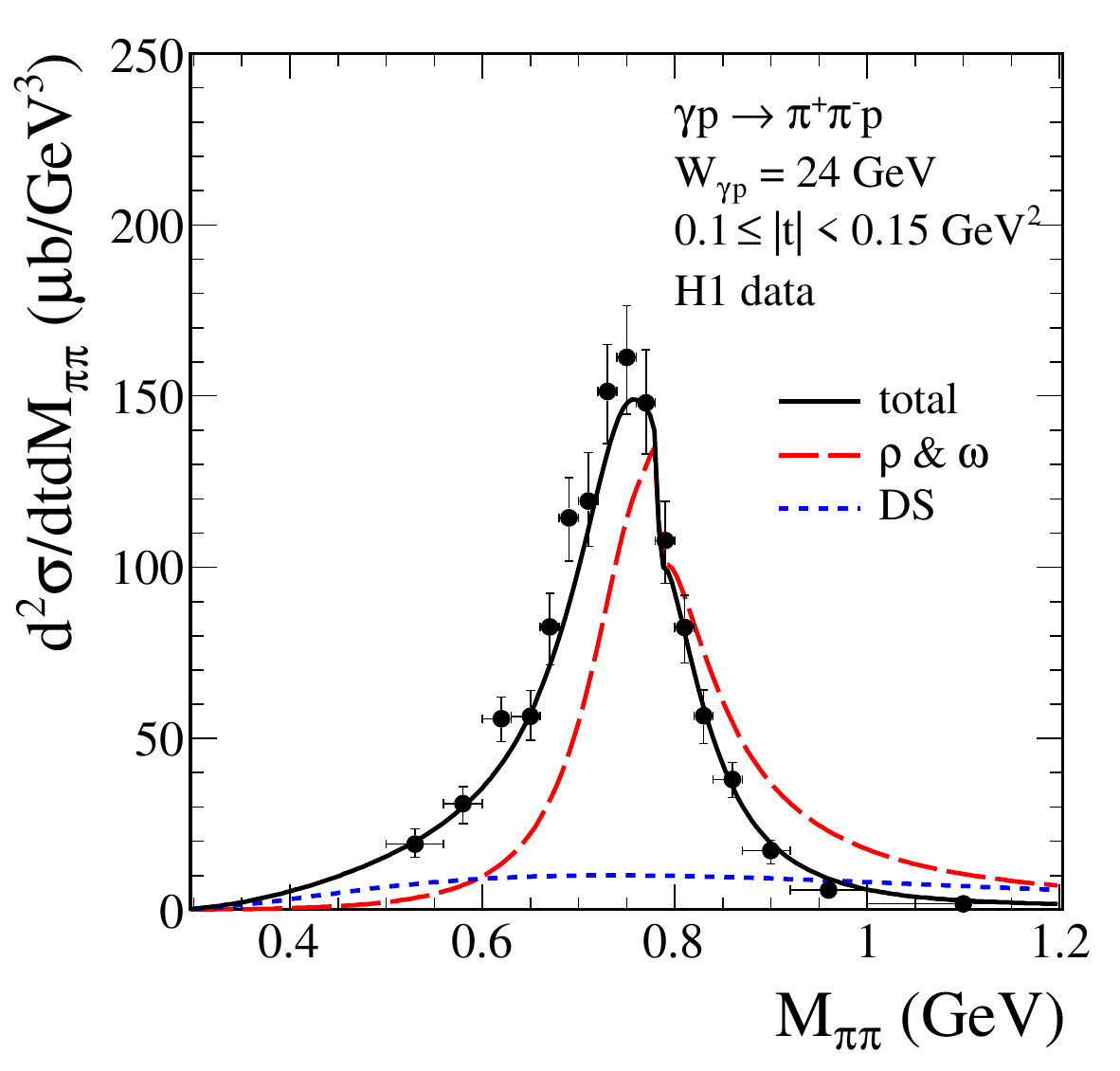}
\includegraphics[width=0.48\textwidth]{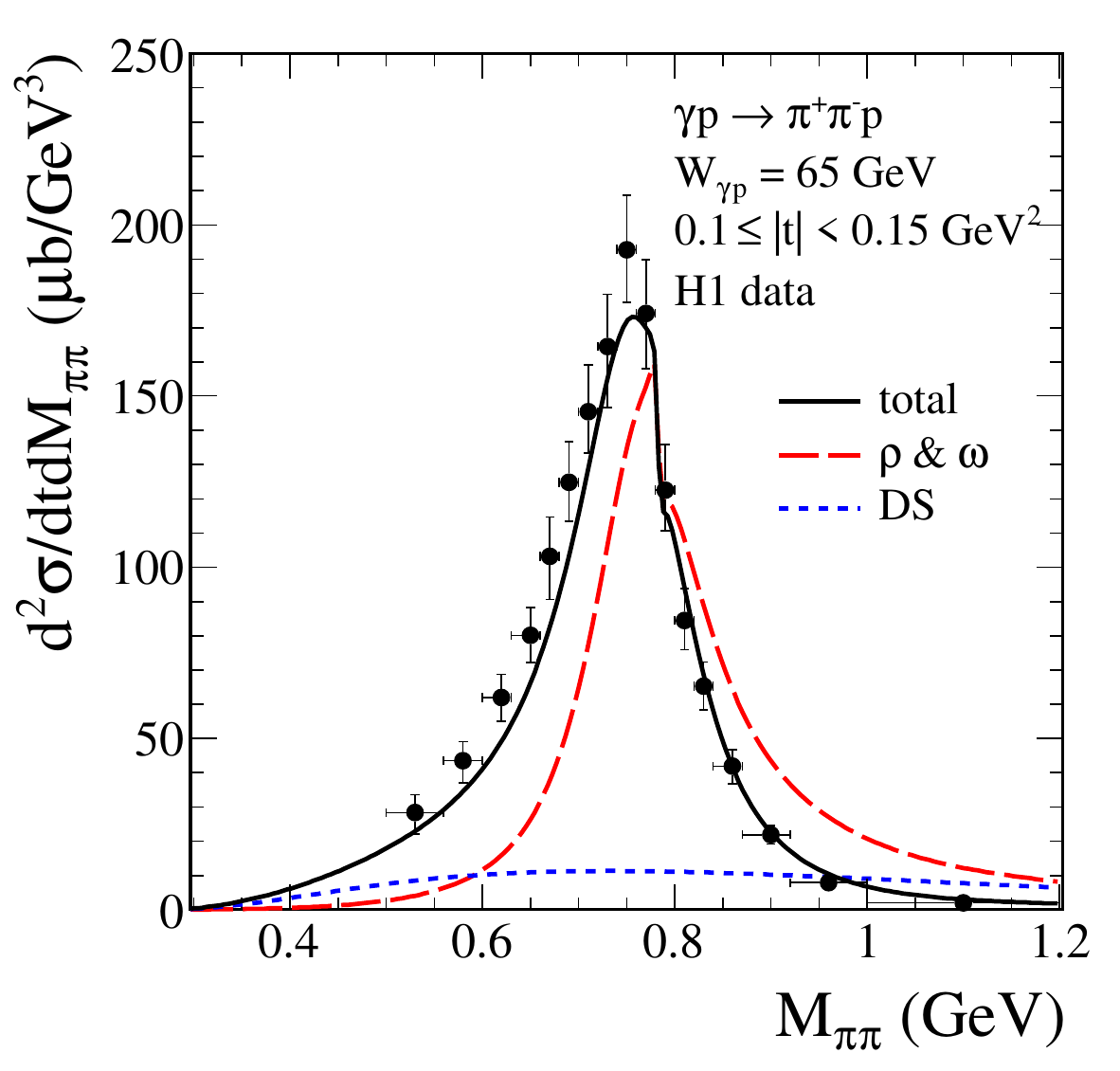}
\includegraphics[width=0.48\textwidth]{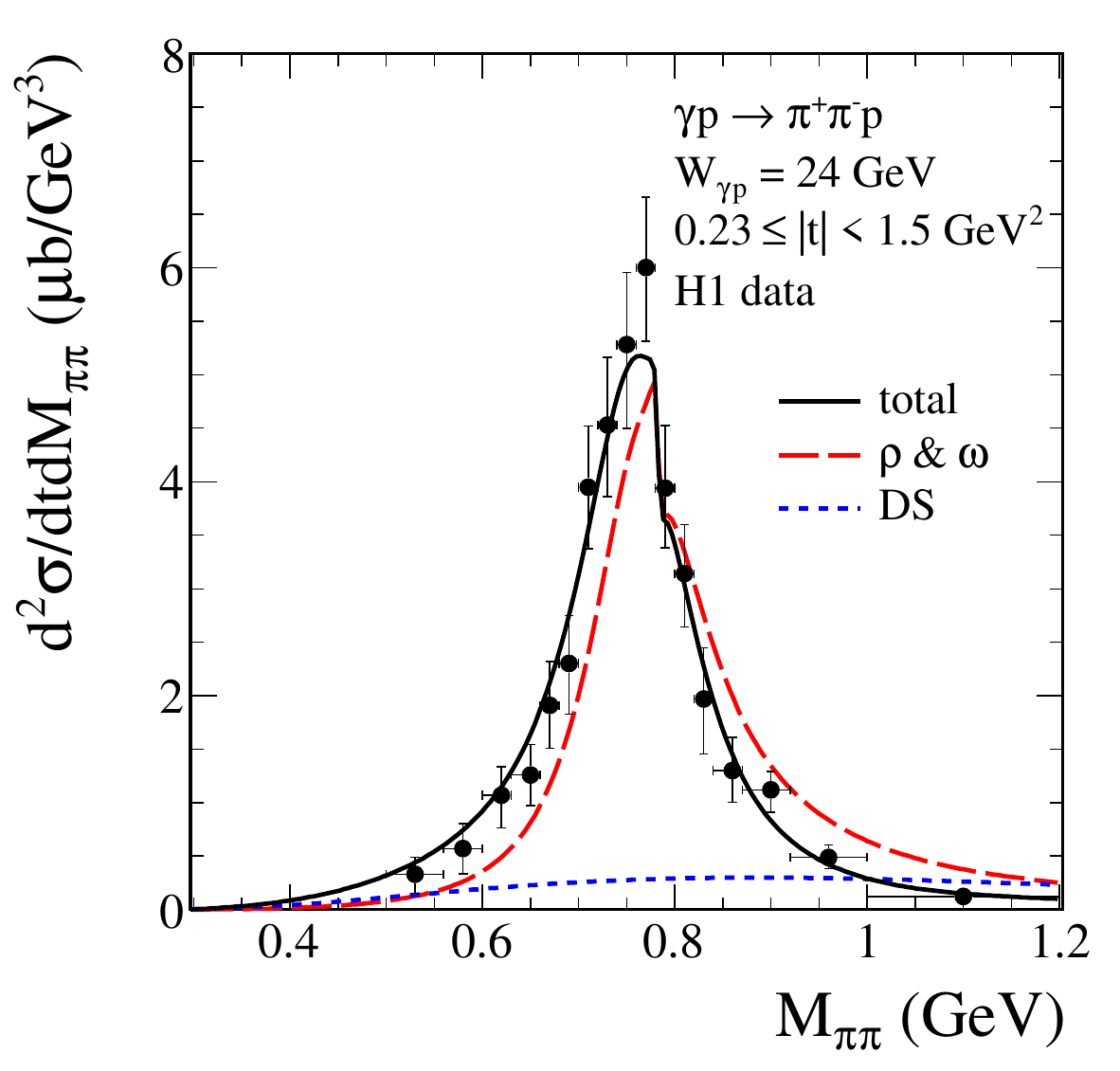}
\includegraphics[width=0.48\textwidth]{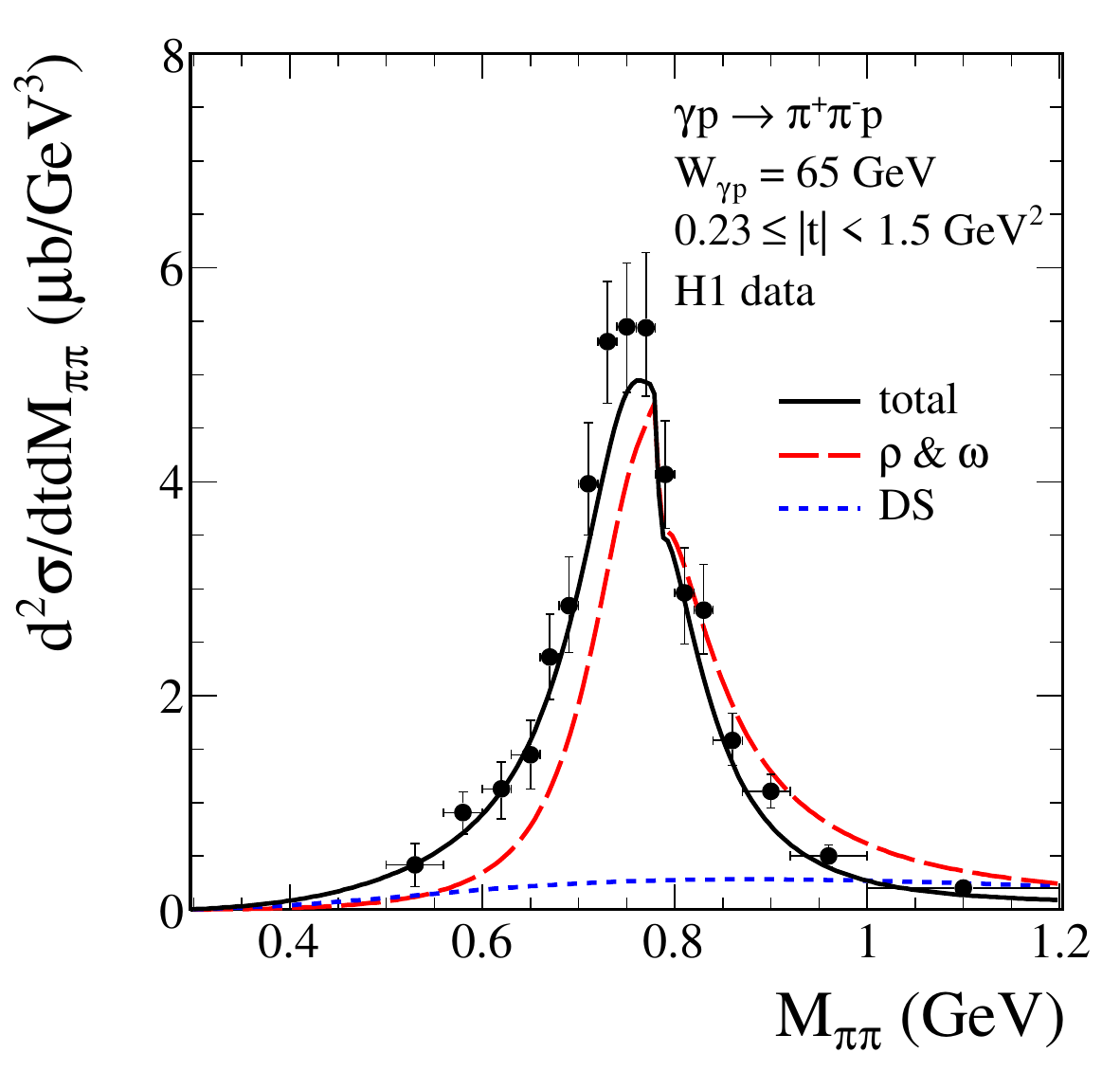}
\caption{\label{fig:3}
The double-differential cross sections $d^{2}\sigma/dtdM_{\pi\pi}$ 
for the reaction $\gamma p \to \pi^{+} \pi^{-} p$
for fixed photon-proton centre-of-mass energies 
$W_{\gamma p} = 24$ and 65~GeV.
The experimental data of the H1 Collaboration are from \cite{H1:2020lzc}
and were measured in the ranges $20 < W_{\gamma p} < 28$~GeV and
$50 < W_{\gamma p} < 80$~GeV
and for different $|t|$ intervals.
The statistical and systematic
uncertainties are shown.
The meaning of the lines is the same as in figure~\ref{fig:1}.
In the calculations we take $\Lambda_{\pi} = 0.8$~GeV, $\Lambda_{V} = 2.0$~GeV, 
and $n_{V} = 2.0$.}
\end{figure}

From figures \ref{fig:2} and \ref{fig:3} we clearly see
that the $\pi^{+}\pi^{-}$ invariant mass distribution is skewed 
and the amount of skewing decreases with increasing $|t|$.

\begin{figure}[tbp]
\centering
\includegraphics[width=0.48\textwidth]{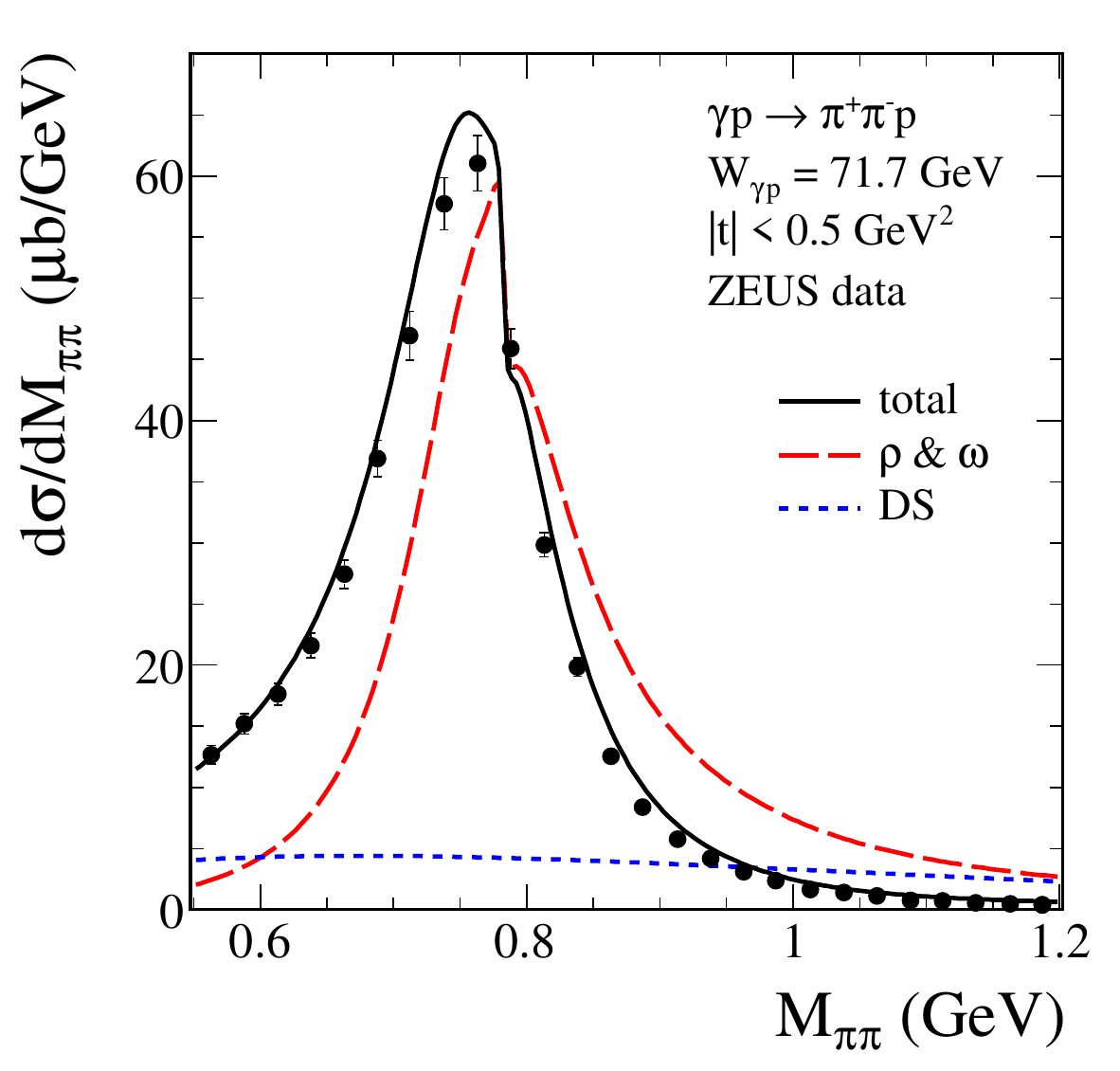}
\includegraphics[width=0.48\textwidth]{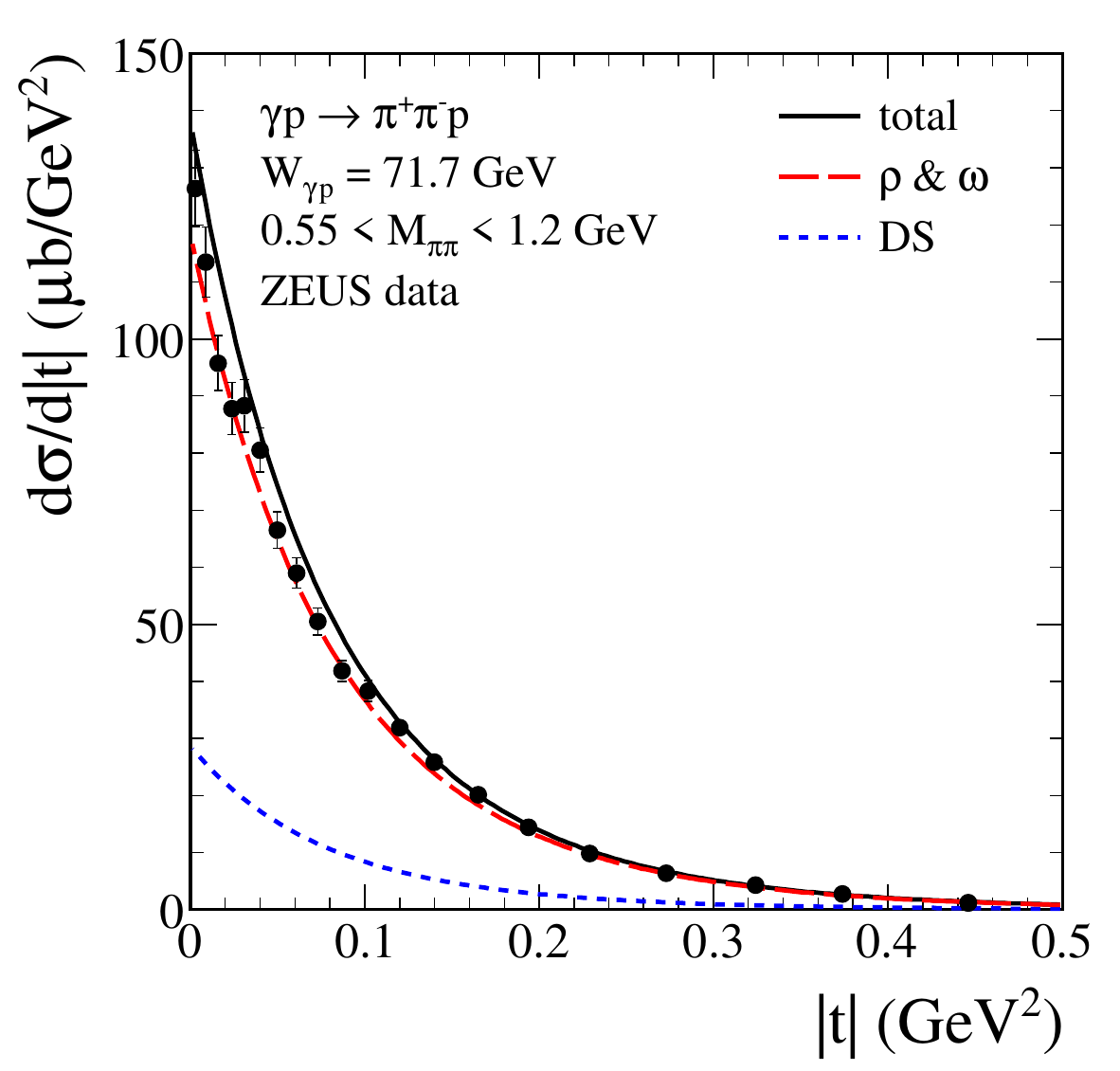}
\caption{\label{fig:4}
The differential distributions 
for the reaction $\gamma p \to \pi^{+} \pi^{-} p$.
The experimental data of the ZEUS Collaboration are from \cite{ZEUS:1997rof}.
Only statistical errors are shown.
The ZEUS data were measured in the kinematic region $50 < W_{\gamma p} < 100$~GeV ($\langle W_{\gamma p} \rangle = 71.7$~GeV) 
and $|t| < 0.5$~GeV$^{2}$.
The calculations are made for fixed $W_{\gamma p} = 71.7$~GeV.
The meaning of the lines is the same as in figure~\ref{fig:1}.
In the calculations we take $\Lambda_{\pi} = 0.8$~GeV, $\Lambda_{V} = 2.0$~GeV, 
and $n_{V} = 2.0$.
}
\end{figure}

\begin{figure}[tbp]
\centering
\includegraphics[width=0.48\textwidth]{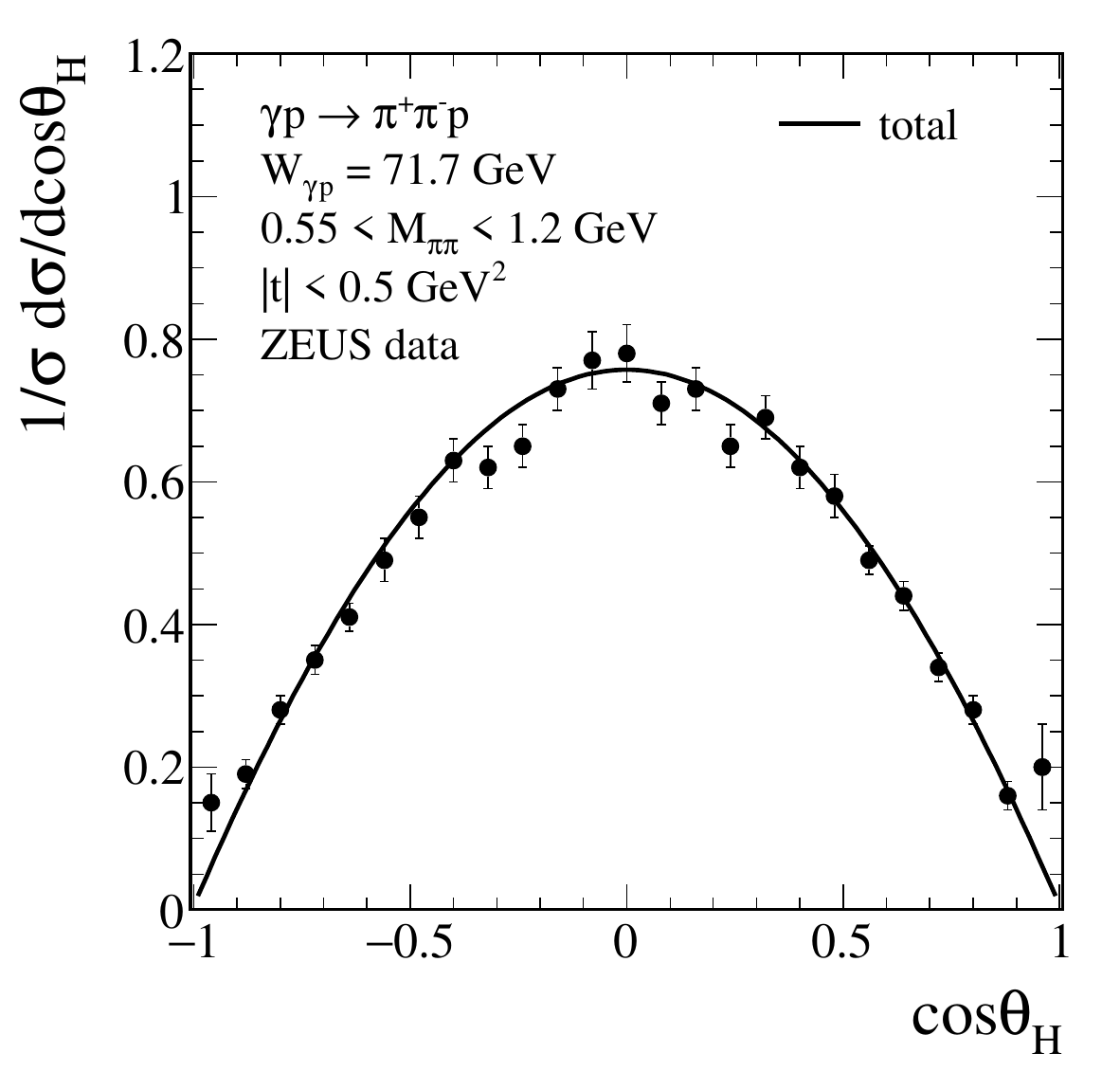}
\includegraphics[width=0.48\textwidth]{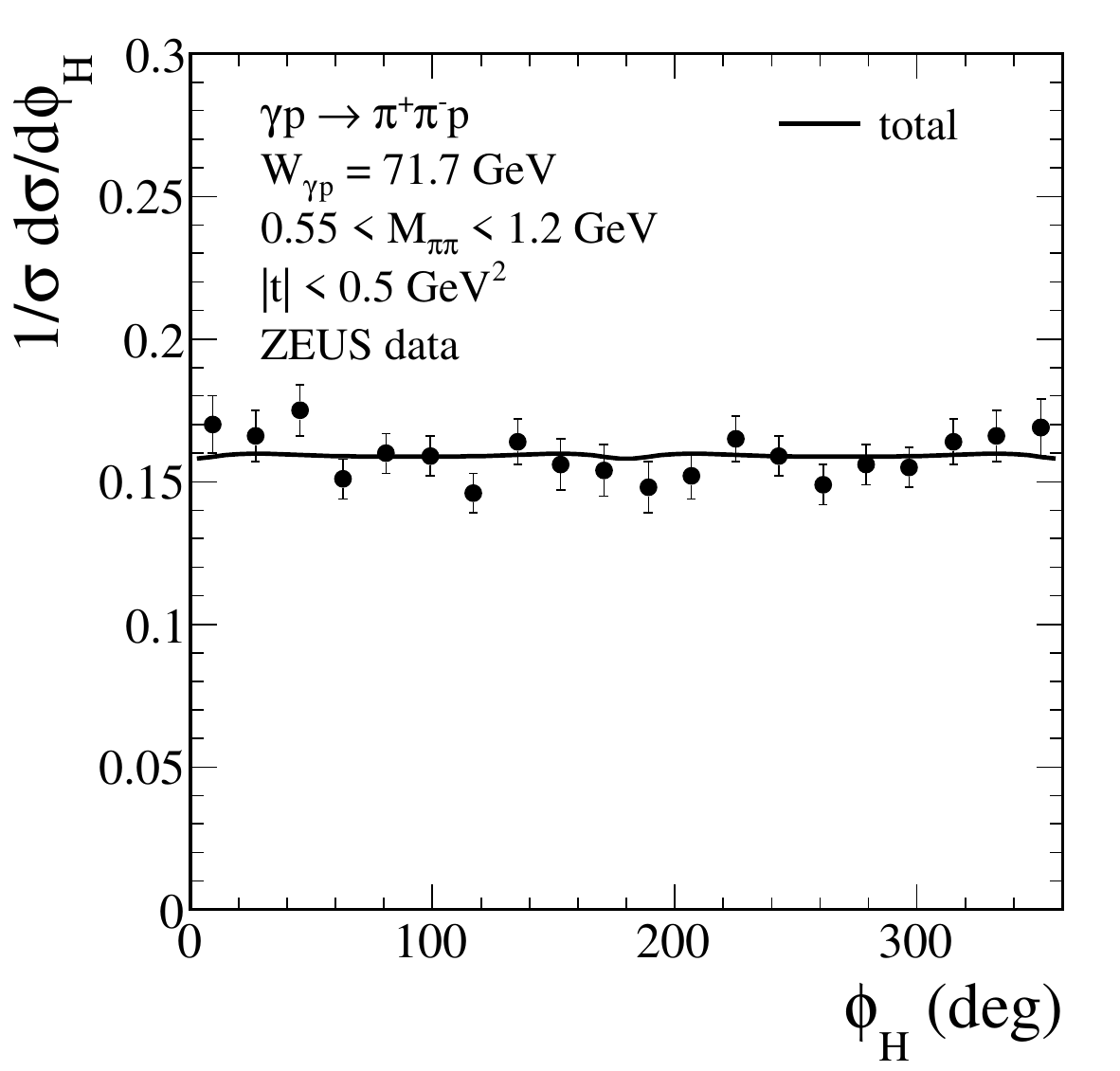}
\caption{\label{fig:5}
The angular distributions of the $\pi^{+}$,
$(1/\sigma) \, d\sigma/d\cos\theta_{H}$
and 
$(1/\sigma) \, d\sigma/d\phi_{H}$,
in the helicity frame.
The full model (total) results are compared with the
experimental data of the ZEUS Collaboration from \cite{ZEUS:1997rof}.
Only statistical errors are shown.
The ZEUS data were measured in the kinematic region 
$50 < W_{\gamma p} < 100$~GeV, 
$|t| < 0.5$~GeV$^{2}$, 
and $0.55 < M_{\pi\pi} < 1.2$~GeV.
Our calculated cross section for $W_{\gamma p} = 71.7$~GeV
is $\sigma = 11.78$~$\mu$b.}
\end{figure}

Figures~\ref{fig:4} and \ref{fig:5} show a comparison
of our model results with the ZEUS data from \cite{ZEUS:1997rof}.
In figure~\ref{fig:5} we show the pion angular distributions
in the rest frame of the $\pi^{+}\pi^{-}$ system.
In \cite{ZEUS:1997rof} the helicity frame as defined in
\cite{Schilling:1969um,Schilling:1973ag} is used.
There we have for the momenta
\begin{eqnarray}
\mathbf{k} = 0\,, \qquad \mathbf{p} + \mathbf{q} = \mathbf{p'} \,,
\label{3.100}
\end{eqnarray}
and the following coordinate unit vectors are chosen as
\begin{align}
\mathbf{e}_{\mathbf{1}}  &=  
- \frac{\mathbf{\hat{q}} - \mathbf{\hat{p}'} (\mathbf{\hat{p}'} \cdot \mathbf{\hat{q}})}{|\mathbf{\hat{p}'} \times \mathbf{\hat{q}}|}
\,, \nn \\
\mathbf{e}_{\mathbf{2}}  &=  
\frac{\mathbf{\hat{p}'} \times \mathbf{\hat{q}}}{|\mathbf{\hat{p}'} \times \mathbf{\hat{q}}|}
\,, \nn \\
\mathbf{e}_{\mathbf{3}}  &= 
\mathbf{-\hat{p}'} \,,
\label{3.101}
\end{align}
with $\mathbf{\hat{p}'} = \mathbf{p}' / |\mathbf{p}'|$,
$\mathbf{\hat{q}} = \mathbf{q} / |\mathbf{q}|$
and $\mathbf{p}'$, $\mathbf{q}$ the three-momenta of the recoil proton
and the $\gamma$ in the $\pi^{+}\pi^{-}$ rest system.
The corresponding polar and azimuthal angles of the $\pi^{+}$
momentum $\mathbf{k_{1}}$ are denoted 
by $\theta_{\rm H}$ and $\phi_{\rm H}$, respectively,
and we have 
$\cos \theta_{\rm H} = -\mathbf{\hat{k}_{1}} \cdot \mathbf{\hat{p}'}$.

For a discussion of other commonly used frames see
appendix~A of \cite{Bolz:2014mya}
and references quoted there.

Note that the helicity frame defined in (A.18) of \cite{Bolz:2014mya} 
by $\mathbf{e}_{\mathbf{i},\mathrm{H}}$
(${\rm i} = 1, 2, 3$)
differs from (\ref{3.101}) by sign changes:
\begin{align}
\mathbf{e}_{\mathbf{1},\mathrm{H}} = -\mathbf{e}_{\mathbf{1}}\,,
\quad 
\mathbf{e}_{\mathbf{2},\mathrm{H}} =  \mathbf{e}_{\mathbf{2}}\,,\quad 
\mathbf{e}_{\mathbf{3},\mathrm{H}} = -\mathbf{e}_{\mathbf{3}}\,.
\label{3.102}
\end{align}
%

\section{Conclusions}
\label{sec:4}

We summarize our findings.

\begin{itemize}

\item
We have in \cite{Lebiedowicz:2025xob} given a calculation of 
the Drell-S{\"o}ding (DS) term which fulfils all requirements of QFT.
In this Addendum, we have updated the DS amplitudes taking into account 
pion off-shell effects in the $\Pom \pi \pi$, $f_{2 \Reg} \pi \pi$,
and $\rho_{\Reg} \pi \pi$ vertices.
We have considered the gauge invariance constraint for these new amplitudes
${\cal N}^{(a,b,c)}_{\mu}$
corresponding to the diagrams of figure~\ref{fig:100}.
Using these constraints we could obtain an exact solution for 
${\cal N}^{(c)}_{\mu}$, given the amplitudes 
${\cal N}^{(a)}_{\mu}$ and ${\cal N}^{(b)}_{\mu}$ in the tensor-pomeron model.
The complete result for our DS amplitude is given in (\ref{2.17a}).
For the vector-meson production ($\rho$, $\omega$) we used the formulas from
section~2.2.1 of \cite{Lebiedowicz:2025xob}.

\item
We have presented results of our model 
for real photoproduction $\gamma p \to \pi^{+} \pi^{-} p$
and compared them with data from the ZEUS \cite{ZEUS:1997rof} 
and H1 \cite{H1:2020lzc} experiments.
As can be seen from figures~\ref{fig:1}--\ref{fig:5} our model
gives a very satisfactory description of the data
for invariant $\pi^{+} \pi^{-}$ masses $M_{\pi \pi} \lesssim 1.2$ GeV.
For a description of the data for $M_{\pi \pi} > 1.2$ GeV
we would certainly have to include contributions of the higher mass mesons,
e.g., $\rho(1450)$, $\rho_{3}(1690)$, and $\rho(1700)$.
This can be done with extensions of the methods explained in
section~2.1 of \cite{Bolz:2014mya}.
From our detailed study of distributions for $M_{\pi \pi} \lesssim 1.2$ GeV
we find from figure~\ref{fig:2} that our tensor-pomeron model,
incorporating both resonant ($\rho^{0}$ and $\omega$) 
and non-resonant (DS) contributions,
is in good agreement with the H1 experimental data.
From figure \ref{fig:3} we see that the $\pi^{+}\pi^{-}$ 
invariant mass distribution is skewed 
and the amount of skewing decreases with increasing $|t|$.
Furthermore, our model with default parameters 
is also in good agreement with the ZEUS data;
see figures \ref{fig:4} and \ref{fig:5}.

\item
In appendix~\ref{sec:A}
we compare our methods and results to those presented
by J.~Pumplin in \cite{Pumplin:1970kp}.

\item
We emphasize that in \cite{Lebiedowicz:2025xob} and the present paper
we are presenting a model for the \textit{complete amplitudes}
of the $\gamma^{(*)} p \to \pi^{+} \pi^{-} p$ reaction
for real and virtual photons with 
$0 \leqslant Q^{2} \lesssim 0.5$~GeV$^{2}$.
Therefore, our model makes predictions for all measurable
quantities in $\pi^{+} \pi^{-}$ photoproduction
and low $Q^{2}$ electroproduction and, thus,
allows for detailed comparison of theory and experiment.

\item
We invite the experimentalists to make a complete and detailed comparison
of our results for the $\gamma p \to \pi^{+} \pi^{-} p$ reaction
to the corresponding data from HERA.
In this way we could get experimental numbers with errors for the (few)
parameters of our model.
These parameters refer to the couplings, intercepts, and slopes of the Regge
trajectories $\Pom$, $f_{2 \Reg}$, $\rho_{\Reg}$, to be compared
with other determinations, and to the shapes of form factors.
Such an analysis would be very useful for the experimental study
of central-exclusive production (CEP) of $\pi^{+}\pi^{-}$ pairs
in ultra-peripheral collisions of proton-proton/ion scattering at the LHC,
and in the future at electron-proton/ion collisions at 
EIC ([63--66] of \cite{Lebiedowicz:2025xob}) 
and LHeC ([67, 68] of \cite{Lebiedowicz:2025xob}).

\item
We think that with the present paper we have made a large step in giving
a QFT solution to a very old problem/puzzle.
Given the shape of the $M_{\pi \pi}$ distribution from the reaction
$e^{+} e^{-} \to \rho^{0}, \omega \to \pi^{+} \pi^{-}$
what is the shape of the corresponding distribution in
$\pi^{+} \pi^{-}$ photoproduction?

\end{itemize}

\appendix

\section{Comparison of our model for the DS contribution with Pumplin's approach}
\label{sec:A}

It is interesting to compare our results for the DS term
for the $\gamma p \to \pi^{+} \pi^{-} p$ reaction
with real photons to Pumplin's ansatz \cite{Pumplin:1970kp}.
This approach was used for instance in \cite{Szczurek:2004xe,
JointPhysicsAnalysisCenter:2024qld}. 

Considering the pomeron exchange only,
Pumplin's formula can be written as
\begin{align}
{\cal M}^{\rm Pumplin}_{\mu, \mathfrak{s'},\mathfrak{s}}(k_{1},k_{2},p',q,p)|_{\Pom} 
= &\;
e 
\left[
\left(
\frac{k_{2 \mu}}{(q, k_{2})} - \frac{(p + p')_{\mu}}{(q, p + p')}
\right) F_{\pi}(t_{\pi,2}) \,
{\cal M}_{\mathfrak{s'},\mathfrak{s}}^{(0,b)}(k_{1},p',q-k_{2},p)|_{\Pom} 
\right. \nonumber\\
& \left.
- 
\left(
\frac{k_{1 \mu}}{(q, k_{1})} - \frac{(p + p')_{\mu}}{(q, p + p')}
\right) F_{\pi}(t_{\pi,1}) \,
{\cal M}_{\mathfrak{s'},\mathfrak{s}}^{(0,a)}(k_{2},p',q-k_{1},p)|_{\Pom}
\right];
\label{A.1}
\end{align}
see eq.~(5) and section~V~A of \cite{Pumplin:1970kp}.

Let us first set the form factors to 1,
\begin{eqnarray}
F_{\pi}(t_{\pi,i}) = 1\,, \quad {\rm for} \;\; i = 1,2\,,
\label{A.100}
\end{eqnarray}
and compare with eqs.~(2.21)--(2.23) of \cite{Lebiedowicz:2025xob}.
Pumplin's ansatz for ${\cal M}^{(c)}_{\mu,\mathfrak{s'},\mathfrak{s}}$
reads then
\begin{align}
{\cal M}^{(c) \, {\rm Pumplin}}_{\mu, \mathfrak{s'},\mathfrak{s}}(k_{1},k_{2},p',q,p)|_{\Pom}
= 
+ e  \frac{(p + p')_{\mu}}{(q, p + p')}
&\left[
{\cal M}_{\mathfrak{s'},\mathfrak{s}}^{(0,a)}(k_{2},p',q-k_{1},p)|_{\Pom} 
\right. \nonumber\\
& \left.
- 
{\cal M}_{\mathfrak{s'},\mathfrak{s}}^{(0,b)}(k_{1},p',q-k_{2},p)|_{\Pom}
\right]\,.
\label{A.101}
\end{align}
Note that Pumplin's assumption is that 
${\cal M}^{(c)}_{\mu}$ is proportional to $(p + p')_{\mu}$.
We also note that a term $(q, p + p')^{-1}$ in ${\cal M}_{\mu}$
which becomes singular for $q \to 0$ would be hard to understand
from the point of view of QFT.

Now we compare (\ref{A.101}) to our result for
${\cal M}^{(c)}_{\mu, \mathfrak{s'},\mathfrak{s}}$ from
(2.23) of \cite{Lebiedowicz:2025xob}:
\begin{eqnarray}
&&{\cal M}_{\mu, \mathfrak{s'},\mathfrak{s}}^{(c)}(k_{1},k_{2},p',q,p)|_{\Pom}
=
2ie 
{\cal F}_{\Pom \pi p}(2 \bar{\nu}, t)
\bigg{\lbrace}
2 \delta_{\mu}^{\;\;\nu} (k_{2} - k_{1}, p' + p)
+ 2 (p'+p)_{\mu} (k_{2} - k_{1})^{\nu}\nonumber\\
&& \quad 
+ (p'+p)_{\mu} (2 - \alpha_{\Pom}(t)) \frac{(p'+p,k_{1}-k_{2})}{16 \bar{\nu}^{2}}\nonumber\\
&& \quad  \times
\bigg{[}
g\left( \frac{2 - \alpha_{\Pom}(t)}{2}, \varkappa \right)
(k_{2} - k_{1} + q)^{\nu}(k_{2}-k_{1}+q,p'+p) \nonumber\\
&& \quad 
+
g\left( \frac{2 - \alpha_{\Pom}(t)}{2}, -\varkappa \right)
(k_{2} - k_{1} - q)^{\nu}(k_{2}-k_{1}-q,p'+p)
\bigg{]}
\bigg{\rbrace} 
\bar{u}_{\mathfrak{s'}}(p') \gamma_{\nu} u_{\mathfrak{s}}(p) \nonumber\\
&& \quad +
2ie {\cal F}_{\Pom \pi p}(2 \bar{\nu}, t)
\bigg{\lbrace}
- 2 (k_{2} - k_{1})_{\mu} 
- (p'+p)_{\mu} \frac{2 - \alpha_{\Pom}(t)}{2} \frac{(p'+p,k_{1}-k_{2})}{16 \bar{\nu}^{2}}\nonumber\\
&& \quad \times
\left[
g\left( \frac{2 - \alpha_{\Pom}(t)}{2}, \varkappa \right)
(k_{2}-k_{1}+q)^{2} +
g\left( \frac{2 - \alpha_{\Pom}(t)}{2},-\varkappa \right)
(k_{2}-k_{1}-q)^{2} 
\right] 
\bigg{\rbrace} \nonumber \\
&& \quad \times
m_{p} \bar{u}_{\mathfrak{s'}}(p') u_{\mathfrak{s}}(p)\,.
\label{A.102}
\end{eqnarray}
The function ${\cal F}_{\Pom \pi p}(2 \bar{\nu}, t)$ 
is defined in (A.52) of \cite{Lebiedowicz:2025xob}.
We see that our result for
${\cal M}^{(c)}_{\mu, \mathfrak{s'},\mathfrak{s}}$
is, in general, \textit{not} proportional to $(p + p')_{\mu}$.
But using the high-energy approximation,
\begin{eqnarray}
\bar{u}_{\mathfrak{s'}}(p') \gamma_{\nu} u_{\mathfrak{s}}(p) &\simeq& (p' + p)_{\nu}\, \delta_{\mathfrak{s'},\mathfrak{s}}\,, \nn \\
2 m_{p} \, \bar{u}_{\mathfrak{s'}}(p') u_{\mathfrak{s}}(p) &\simeq& 0\,,
\label{A.2} 
\end{eqnarray}
we find from (2.21) and (2.22) of \cite{Lebiedowicz:2025xob}
\begin{eqnarray}
{\cal M}_{\mathfrak{s'},\mathfrak{s}}^{(0, \,a)}(k_{2},p',q-k_{1},p)|_{\Pom} 
&\simeq &
i{\cal F}_{\Pom \pi p}(2 \bar{\nu}, t)
\left[
1 + (2 - \alpha_{\Pom}(t)) \frac{\varkappa}{2} \,
g\left( \frac{2 - \alpha_{\Pom}(t)}{2}, \varkappa \right)
\right] \nonumber\\
&& \times
2(k_{2}-k_{1}+q, p'+p)^{2} \,
\delta_{\mathfrak{s'},\mathfrak{s}} \,,
\label{A.3} \\
{\cal M}_{\mathfrak{s'},\mathfrak{s}}^{(0,\,b)}(k_{1},p',q-k_{2},p)|_{\Pom}&\simeq &
i{\cal F}_{\Pom \pi p}(2 \bar{\nu}, t)
\left[
1 - (2 - \alpha_{\Pom}(t)) \frac{\varkappa}{2} \,
g\left( \frac{2 - \alpha_{\Pom}(t)}{2}, -\varkappa \right)
\right] \nonumber\\
&& \times
2(k_{2}-k_{1}-q, p'+p)^{2} \,
\delta_{\mathfrak{s'},\mathfrak{s}}\,.
\label{A.4}
\end{eqnarray}
Here we have from (A.3)--(A.5) of \cite{Lebiedowicz:2025xob}
\begin{eqnarray}
\nu_{1} &=& \frac{1}{4}\left[ (p+p',k_{1}-k_{2}) + (p+p',q) \right]\,, \nn\\
\nu_{2} &=& \frac{1}{4}\left[ (p+p',k_{2}-k_{1}) + (p+p',q) \right]\,,\nn\\
\bar{\nu} &=& \sqrt{\frac{1}{2}(\nu_{1}^{2} + \nu_{2}^{2})}\,, \nn\\
\varkappa &=& \frac{2(q,p+p')(p+p',k_{1}-k_{2})}{16 \bar{\nu}^{2}}\,.
\label{A.5}
\end{eqnarray}
Furthermore, we have
\begin{eqnarray}
(q, p + p') = \frac{s-u}{2} = 2 \nu_{1} + 2 \nu_{2}\,,
\label{Pumplin_aux}
\end{eqnarray}
where
\begin{eqnarray}
s & \equiv & W_{\gamma p}^{2} = (q + p)^{2} = (k_{1} + k_{2} + p')^{2}\,, \nn\\
u &=& (q - p')^{2} = (p - k_{1} - k_{2})^{2}\,.
\end{eqnarray}

\begin{figure}[tbp]
\centering
(a)\includegraphics[width=0.46\textwidth]{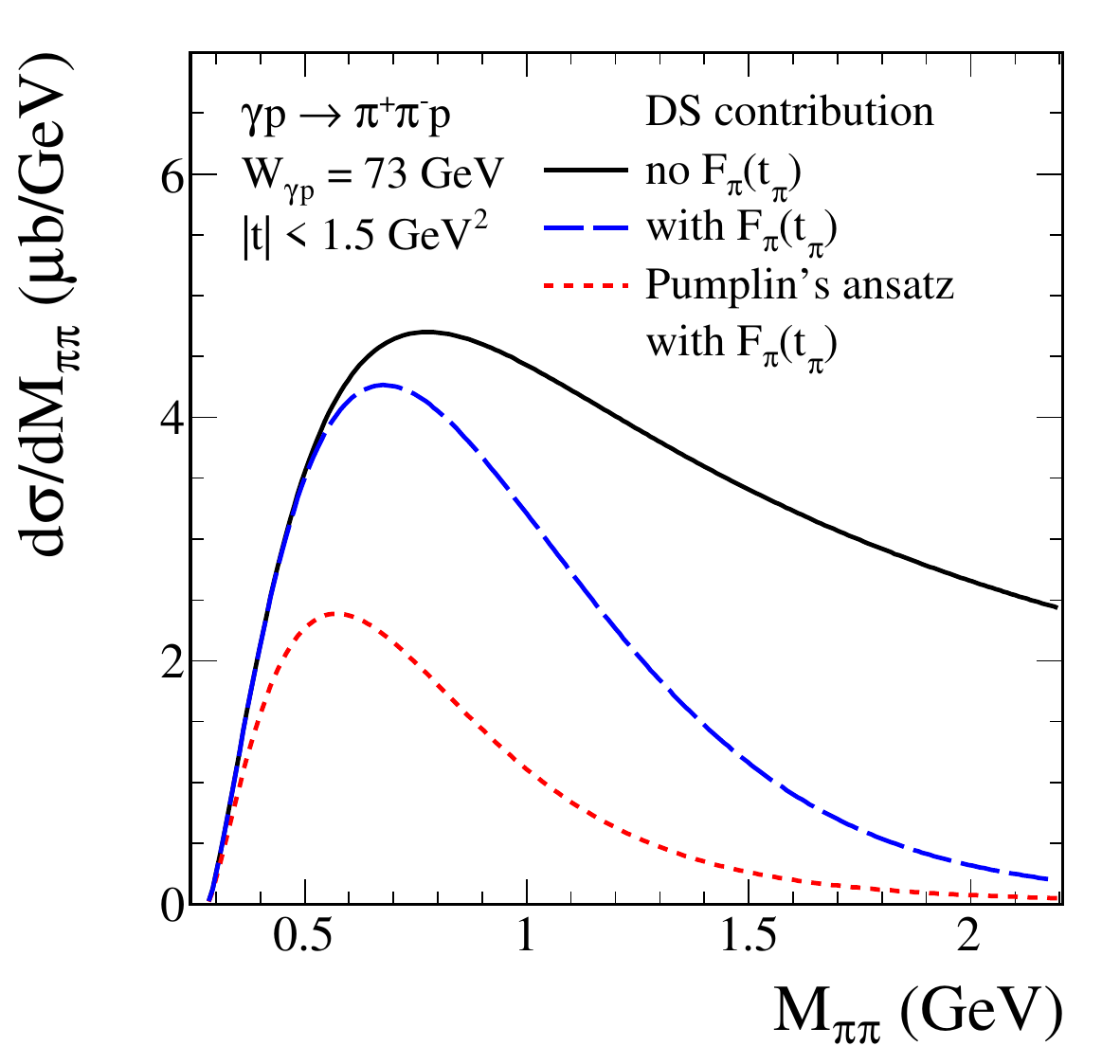}
(b)\includegraphics[width=0.46\textwidth]{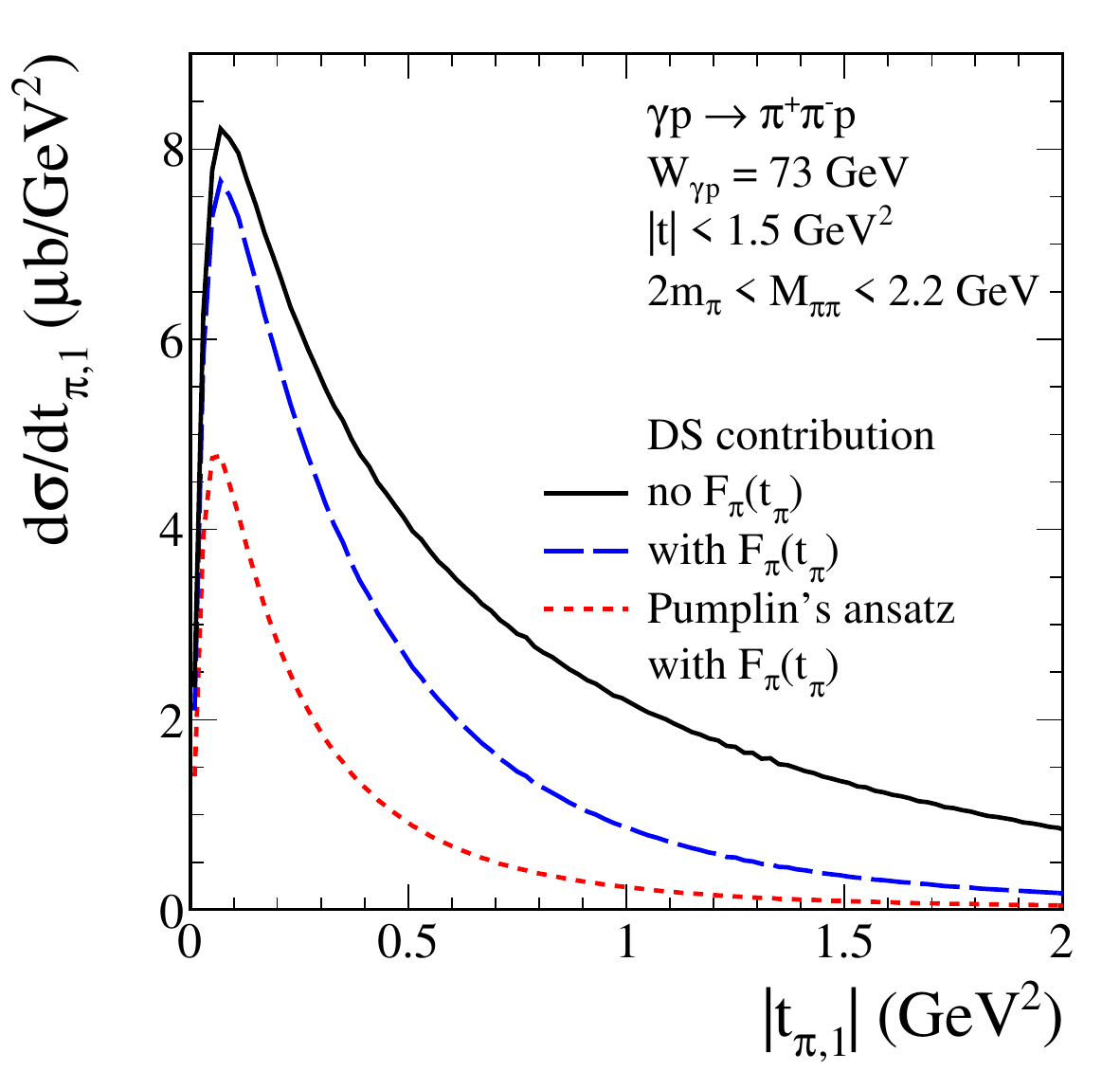}
\caption{\label{fig:DS_comparison}
Comparison of our DS model (\ref{A17}) and Pumplin's ansatz (\ref{A.1}).
In (a) we show the $M_{\pi\pi}$, 
in (b) the $|t_{\pi,1}|$ distributions.
The $|t_{\pi,2}|$ distributions are identical to those of $|t_{\pi,1}|$.
In both approaches the same type of form factor
$F_{\pi}(t_{\pi,i})$ for $i = 1,2$ (\ref{A21}) was assumed.
Here we take $\Lambda_{\pi} = 0.8$~GeV.
The black solid lines represents the results 
for $F_{\pi}(t_{\pi,i}) = 1$ in (\ref{A19}).}
\end{figure}

In the high-energy approximation (\ref{A.2})
${\cal M}_{\mu, \mathfrak{s'},\mathfrak{s}}^{(c)}$ from (\ref{A.102})
indeed becomes proportional to $(p + p')_{\mu}$,
and the unwanted term $(p' + p,q)^{-1}$ drops out:
\begin{eqnarray}
{\cal M}_{\mu, \mathfrak{s'},\mathfrak{s}}^{(c)}(k_{1},k_{2},p',q,p)|_{\Pom}
&\simeq & (p'+p)_{\mu}\, ie \,
{\cal F}_{\Pom \pi p}(2 \bar{\nu}, t) \, \delta_{\mathfrak{s'},\mathfrak{s}}
\nn \\
&& \times \bigg{\lbrace}  8 (k_{2} - k_{1}, p' + p)
+ (2 - \alpha_{\Pom}(t)) \frac{(p'+p,k_{1}-k_{2})}{16 \bar{\nu}^{2}}\nonumber\\
&& \times
\bigg{[}
g\left( \frac{2 - \alpha_{\Pom}(t)}{2}, \varkappa \right)
2(k_{2} - k_{1} + q,p'+p)^{2} \nonumber\\
&& + 
g\left( \frac{2 - \alpha_{\Pom}(t)}{2}, -\varkappa \right)
2(k_{2} - k_{1} - q,p'+p)^{2}
\bigg{]}
\bigg{\rbrace} \nn \\
&\simeq & \frac{(p'+p)_{\mu}}{(p'+p,q)} \,e \,
\bigg{[}
{\cal M}_{\mathfrak{s'},\mathfrak{s}}^{(0, \,a)}|_{\Pom}
-
{\cal M}_{\mathfrak{s'},\mathfrak{s}}^{(0, \,b)}|_{\Pom}
\bigg{]}
\,.
\label{A.6}
\end{eqnarray}
Thus, Pumplin's ansatz works as an \textit{approximate relation}
in the high-energy limit if the form factors
are set to 1; see (\ref{A.100}).
We have checked this numerically; see figure~\ref{fig:DS_comparison}.

But now we come to the formulas including non trivial form factors 
$F_{\pi}(t_{\pi,i})$.
Then we have to compare Pumplin's formula (\ref{A.1})
with our formulas
${\cal N}^{(a)}_{\mu, \mathfrak{s'},\mathfrak{s}}|_{\Pom}$,
${\cal N}^{(b)}_{\mu, \mathfrak{s'},\mathfrak{s}}|_{\Pom}$,
and
${\cal N}^{(c)}_{\mu, \mathfrak{s'},\mathfrak{s}}|_{\Pom}$
from eqs.~(\ref{A9}), (\ref{A11}), and (\ref{A16}).
We see from (\ref{A16}) that ${\cal N}^{(c)}_{\mu, \mathfrak{s'},\mathfrak{s}}|_{\Pom}$ 
has a term proportional to $(k_{2} - k_{1})_{\mu}$,
and this term is \textit{not} small at high energies.
This is verified by the numerical calculations.

In figure~\ref{fig:DS_comparison} we show 
the effect of the off-shellness of the pions
on the $M_{\pi \pi}$ 
and $|t_{\pi,1}|$ distributions
calculated at $W_{\gamma p} = 73$~GeV and for $|t| < 1.5$~GeV$^{2}$.
Plotted are the results for the DS contribution 
for real photons ($q^{2} = 0$)
taking only pomeron exchange into account 
and using the small-angle high-energy approximation.
Results for our new model and Pumplin's ansatz obtained 
with the same off-shell pion form factors $F_{\pi}(t_{\pi,i})$ (\ref{A21})
with $\Lambda_{\pi} = 0.8$~GeV are shown.
For comparison, 
we show the result corresponding to $F_{\pi}(t_{\pi,i}) = 1$,
see the black solid line.
When off-shell pion effects are neglected, 
both approaches give very similar results within the assumptions made.
But with form factors our results and those of Pumplin's ansatz
are very different.

\acknowledgments
We thank Stefan Schmitt for useful discussions.



\end{document}